\documentclass[prd,preprint,eqsecnum,nofootinbib,amsmath,amssymb,
               tightenlines]{revtex4}

\usepackage{graphicx}
\usepackage{bm}
\usepackage{verbatim}

\def\p{{\bf p}}

\def\E{{\bf E}}

\def\N{{\mathcal N}}

\def\st{\begin{equation}}
\def\stp{\end{equation}}
\def\bg{\begin{eqnarray}}
\def\nd{\end{eqnarray}}
\def\Eq#1{Eq.~(\ref{#1})}

\def\Sect#1{Section~\ref{#1}}

\def\drangle{\rangle\!\rangle}
\def\dlangle{\langle\!\langle}

\def\llangle{\left\langle}
\def\rrangle{\right\rangle}

\def\N{\mathcal{N}}
\def\oh{\slantfrac{1}{2}}

\def\gsim{\mbox{~{\protect\raisebox{0.4ex}{$>$}}\hspace{-1.1em}
	{\protect\raisebox{-0.6ex}{$\sim$}}~}}

\def\nott#1{\setbox0=\hbox{$#1$}                
   \dimen0=\wd0                                 
   \setbox1=\hbox{/} \dimen1=\wd1               
   \ifdim\dimen0>\dimen1                        
      \rlap{\hbox to \dimen0{\hfil/\hfil}}      
      #1                                        
   \else                                        
      \rlap{\hbox to \dimen1{\hfil$#1$\hfil}}   
      /                                         
   \fi}                                         %

\advance\parskip 1.9pt
\advance\voffset -0.2in

\def\st{\begin{equation}}
\def\stp{\end{equation}}
\def\bg{\begin{eqnarray}}
\def\nd{\end{eqnarray}}

\begin{document}

\vspace*{1cm}

\title{Simulating elliptic flow with viscous hydrodynamics}
\author{K. Dusling}
\author{D. Teaney}
\affiliation{Department of Physics \& Astronomy, State University of New York, Stony Brook, NY 11794-3800, U.S.A.}
\date{\today}

\begin{abstract}
In this work we simulate a viscous hydrodynamical model of non-central
Au-Au collisions in 2+1 dimensions, assuming longitudinal boost
invariance.  The model fluid equations were proposed by \"{O}ttinger 
and Grmela \cite{OG}.  Freezeout is signaled when the viscous corrections
become large relative to the ideal terms. Then viscous corrections to
the transverse momentum and differential elliptic flow spectra are
calculated. When viscous corrections to the thermal distribution
function are not included, the effects of viscosity on elliptic flow
are modest.  However, when these corrections are included, the elliptic
flow is strongly modified at large $p_T$. We also investigate the
stability of the viscous results by comparing the non-ideal  components
of the stress tensor ($\pi^{ij}$) and their influence on the $v_2$
spectrum to the expectation of the Navier-Stokes equations ($\pi^{ij}
= -\eta \llangle \partial_i u_j \rrangle$).  We argue that when the
stress tensor deviates from the Navier-Stokes form the dissipative
corrections to spectra are too large for a hydrodynamic description to
be reliable.  For typical RHIC initial conditions this happens for
$\eta/s \gsim 0.3$.  
\end{abstract}

\maketitle
\setcounter{tocdepth}{2}

\section{Introduction}
\subsection{Motivation}
One of the first and most exciting observations from the 
Relativistic Heavy Ion Collider (RHIC) was the very strong 
elliptic flow in non-central collisions \cite{Adams:2005dq,Adcox:2004mh}. 
The elliptic flow is quantified by the anisotropy of
particle production with respect to the 
reaction plane $v_2$
\st
      v_2 = \llangle \frac{p_x^2 - p_y^2}{p_x^2 + p_y^2} \rrangle,
\stp
and can be measured as a function of $p_T$, rapidity, centrality 
and particle type. 

The adopted interpretation of the $v_2$ measurements is that the medium
responds as a fluid to the differences in pressure gradients in the $x$
and $y$ directions. The fluid then  expands preferentially in the
reaction plane and establishes the observed momentum space anisotropy.
This hydrodynamic interpretation is supported by the qualified success
of ideal hydrodynamic models in describing a large variety of data over
a range of colliding systems and energies \cite{Hirano:2004en,Teaney:2001av,Kolb:2000fh,Huovinen:2001cy,Nonaka:2006yn}.  Nevertheless, the
hydrodynamic interpretation of the flow results is not unassailable.  A
back of the envelope estimate of viscous corrections to hydrodynamic
results \cite{Danielewicz:1984ww} suggests that viscous corrections are actually rather
large, {\em i.e.,} the mean free path is comparable to the system size
\cite{Drescher:2007cd}.
 These estimates are best conveyed in terms of the shear
viscosity to entropy ratio, $\eta/s$.  The conditions for partial
equilibrium at RHIC are so unfavorable that at unless $\eta/s$ is small
(say $0.5$ or less), it is difficult to imagine that the medium would
participate in a coordinated  collective flow.

From a theoretical perspective, it is difficult to  reliably estimate
$\eta/s$ in the vicinity of the QCD phase transition where the system
is strongly coupled.  Lattice QCD measurements of transport are hard
(perhaps impossible \cite{Petreczky:2005nh,Aarts:2002cc}) though recent efforts have lead to estimates which
are not incompatible with the hydrodynamic interpretation of RHIC
results \cite{Aarts:2007wj,Meyer:2007ic}.  In a strict perturbative setting (where the
quasi-particle picture is exact) $\eta/s$ is large $\sim 1/g^4$.
Nevertheless an extrapolation of weak coupling results to moderate
coupling also leads to an $\eta/s$ which is perhaps reconcilable with
the hydrodynamic interpretation \cite{Arnold:2003zc,Baymetal}.  
Finally, these perturbative estimates
should be contrasted with $\N=4$ Super Yang Mills at strong coupling,
where $\eta/s$ is $1/4\pi$ \cite{Policastro:2001yc,Kovtun:2004de}.  Although $\N=4$  SYM is not QCD,  the
calculation was important because it showed that there is at least one theory where $\eta/s$ is
sufficiently small that collective phenomena would be observed under
conditions similar to those produced at RHIC.

From a phenomenological perspective one of the most compelling evidences
for the hydrodynamic interpretation of RHIC flow results is the fact
that the deviations  from hydrodynamics are qualitatively reproduced by
kinetic theory \cite{Molnar:2001ux,Xu:2004mz}. In particular, kinetic theory calculations generically
reproduce the flattening of $v_2(p_T)$ at higher $p_T$, and the
reduction of elliptic flow at large impact parameters.  Some aspects of
these kinetic theory results can be understood by considering the first
viscous corrections to the thermal distribution function \cite{Teaney:2003kp}. These
estimates  motivated full viscous hydrodynamic simulations of the
elliptic flow which will be performed in this work.  Recently such
viscous simulations have been performed by two other groups \cite{Heinzv2,RRv2} and
we will compare our results to these works in \Sect{sec:compare}.  A brief
discussion of the history surrounding viscous relativistic
hydrodynamics is given below.

\subsection{Viscous Hydrodynamics}
\label{viscoushydro}

The Navier-Stokes equations describe viscous corrections to ideal fluid
flow by keeping  terms up to first order in gradients of ideal quantities \cite{LL}.
The resulting equations are parabolic which permit acausal signal propagation \cite{Lindblom:1985}.  For instance, the stress tensor
instantaneously adjusts to any thermodynamic force, $\partial_iu_j$.  
This is, of
course, an unphysical picture  since the stress tensor should relax 
to the  thermodynamic forces over a typical collision timescale.

One would therefore like a phenomenological theory that explains this
relaxation correctly.  Much work has been done in this direction but
there is still no completely satisfactory theory.  
Probably the most used model is that of 
Israel and Stewart \cite{IS,Israel:1979wp}, but there are also others by Lindblom and
Geroch \cite{LG}, Pav\'on, Jou and Casas-V\'asquez \cite{PJC} and also
by \"{O}ttinger and Grmela \cite{OG,Ottinger} which is used in this work. In fact a wide class of models was developed by Lindblom and Geroch in two separate papers \cite{LG,LG2} 

All of the above theories have the same behavior: they relax on small
time scales to the first-order relativistic Navier-Stokes equations
and have some generalized entropy which increases as a function of
time.
It was shown by Lindblom \cite{Lindblom} that for a large class 
class of these second order theories, the physical fields should be
indistinguishable from the simple Navier-Stokes form.  To paraphrase
Lindblom; any measurement of the stress energy tensor or particle
current on a time scale larger than the microscopic time scale will be
indistinguishable from the Navier-Stokes theory.  
The differences between the causal theories 
and the acausal Navier-Stokes equations 
are indicative of the corrections quantitatively captured by the full kinetic theory.
Nevertheless, the causal theories provide
a qualitative guide to the magnitude of these corrections 
 \cite{Forster}.  However, the form of these corrections implicitly 
assumes a good quasi-particle description which may not exist in
a strongly coupled plasma \cite{Teaney:2006nc}.

There has been a large body of work in applying dissipative theories to
central heavy-ion collisions \cite{Muronga,BRW,Teaney:2003kp}.  
Perhaps a particle method will ultimately be the best way to include the 
effects of viscosity and the corresponding fluctuations in the stress tensor \cite{Koide:2006ef,Espanol,Gavin:2006xd}. 
Even though the equations for non-central (2+1 dimensions) dissipative hydrodynamics are known ({\em e.g.} \cite{HSC}), only recently
have results come out which simulate non-central heavy ion collisions
\cite{Heinzv2, RRv2, Ch}.  Further discussion of these results will be  
given in the discussion.

\section{The Hydrodynamic Model}

In the following section we outline the equations of motion for the
hydrodynamical model used in the following simulations.  We start by
summarizing the well known first-order Navier-Stokes theory.  Then we
outline the equations required for a second-order causal description of
dissipative fluid dynamics.  This is done assuming a boost invariant
expansion as first proposed by Bjorken \cite{Bjorken:1982qr}, where the equations
of motion are expressed in terms of the proper time
$\tau=\sqrt{t^2-z^2}$ and the spatial rapidity
$\eta_s=\frac{1}{2}\text{ln}\frac{t+z}{t-z}$. The cartesian coordinate z
denotes the position along the beam axis while x,y label positions
transverse to the beam axis.

\subsection{1$^{st}$ Order Viscous Hydrodynamics - Navier Stokes}

Viscous hydrodynamics was originally outlined in the first-order
Navier-Stokes approximation where the energy momentum tensor and baryon
flux is a sum of their ideal and dissipative parts:
\bg
T^{\mu\nu} &=& \epsilon u^{\mu} u^{\nu} + (p + \Pi) \Delta^{\mu\nu} + \pi^{\mu\nu}\, , \\
n^\mu&=&n u^\mu+j_d^\mu\, ,
\nd
where $p, \epsilon, n$ and $u^\mu=(\gamma,\gamma{\bf \text{v}})$ are
the pressure, energy density, baryon density and four-velocity of the
fluid.  We use the convention that
$g^{\mu\nu}=\text{diag}(-1,+1,+1,+1)$ and therefore $u^\mu u_\mu=-1$.
The dissipative terms, $\pi$ and $j_d$ depend on the definition of the
local rest frame (LRF) of the fluid.  A specific form of $\pi^{\mu\nu}$
and $v^\mu$ can be found using the Landau-Lifshitz definition \cite{LL}
of the LRF ($u_\mu \pi^{\mu\nu}=0$), constraining the the entropy to
increase with time and by working within the Navier-Stokes
approximation (keeping terms to first order in gradients only) resulting in
\bg
\pi^{\mu\nu}&=&-\eta(\nabla^\mu u^\nu+\nabla^\nu u^\mu-\frac{2}{3}\Delta^{\mu\nu}\nabla_\beta u^\beta)\, , \\
\Pi&=&-\zeta \nabla_\beta u^\beta\, , \\
j_d^\mu&=&-\kappa(\frac{nT}{\epsilon+p})^2\nabla^\mu(\frac{\mu}{T})\, ,
\label{eq:pimunu}
\nd
where $\kappa, \eta$ and $\zeta$ are the heat conduction, shear and bulk
viscosities of the fluid with temperature $T$ and chemical potential
$\mu$.  The viscous tensor is constructed with the differential
operator $\nabla^\mu=\Delta^{\mu\nu}d_\nu$ where
$\Delta^{\mu\nu}=g^{\mu\nu}+u^\mu u^\nu$ is the local three-frame
projector and $d_\mu u^\nu=\partial_\mu
u^\nu+\Gamma^\nu_{\gamma\mu}u^\gamma$ is the covariant derivative.  

The transport coefficients in a quark-gluon plasma and also in the
hadronic gas were studied in Refs.~\cite{Danielewicz:1984ww,Prakash:1993bt,Baymetal,Arnold:2003zc}.  It was found that the dominate
dissipative mechanism was shear viscosity in both the QGP and hadronic
gas. Bulk viscosity may however dominate in the transition region 
\cite{Kharzeev:2007wb}.  
Heat transport can be ignored in the limit that $\mu_B\ll T$ which
is the limit taken here.  
   
In the following work we will consider viscous effects in a quark-gluon
plasma phase only.  
For this purpose we  
consider a constant shear to entropy ratio, $\eta/s=\mbox{const}$ 
and a massless gas $p=1/3 \epsilon$. Future work will discuss viscosity in the mixed
and hadronic phases.  From this point on we will neglect
the thermal conductivity.  We keep the bulk viscosity in the
equations for consistency, but always set $\zeta=0$ in any
calculations. 

\subsection{2$^{nd}$ order Viscous Hydrodynamics}
\label{GenericSect}

In order to render a second order theory it is necessary to introduce
additional  variables.  These variables will relax on very short time
scales to the standard thermodynamic quantities in the first order
theory, but an evolution equation for them is still required in order
to avoid acausal signal propagation.  One such theory that has been
used in a number of works was introduced by Israel and Stewart
\cite{IS}.  Instead we use a theory developed by \cite{OG,Ottinger} due
to its appealing structure when implemented numerically.  However, as discussed above, all
of these theories should agree ({\em i.e.,} they all relax on short
time scales to the same the first-order equations).  

We now summarize the evolution equations used in the current analysis
following the mathematical structure outlined in Ref.~\cite{Ottinger}.  We use a
simplified version of the model for deviations of the stress energy
tensor close to equilibrium.  The new dynamical variable that is
introduced is the tensor variable $c_{\mu\nu}$ which will later be
shown to be closely related to the velocity gradient tensor,
$\pi_{\mu\nu}$.   The tensor variable $c_{\mu\nu}$ is conveniently
defined to have the property
\bg
c_{\mu\nu}u^\nu=u_\mu\, ,
\nd 
and the energy momentum tensor is given by
\bg
T^{\mu\nu}=(\epsilon-u_\alpha \mathbb{P}^{\alpha\beta}u_\beta)u^\mu u^\nu+\mathbb{P}^{\mu\nu}\, .
\label{eq:Tmunu}
\nd
The explicit form of the stress tensor $\mathbb{P}^{\mu\nu}$ is given
in \cite{Ottinger} and has a fairly complicated form.  The discussion
in simplified by by considering small deviations from local thermal
equilibrium and working in the local rest frame where the stress tensor
is approximated as
\bg
T^{ij}_{LRF}=p(\delta^{ij}-\alpha c^{ij})\, ,
\nd
where $\alpha$ is a small parameter related to the relaxation time (see
appendix \ref{sec:a}).  The equations of motion are dictated by
conservation of energy and momentum which is given by $d_\mu T^{\mu\nu}=0$.
In addition an evolution equation for the generalized mechanical force
tensor is also needed and is given by \cite{Ottinger}
\bg
u^\lambda(\partial_\lambda c_{\mu\nu}-\partial_\mu c_{\lambda\nu}-\partial_\nu c_{\mu\lambda})=\frac{-1}{\tau_0}\overline{c}_{\mu\nu}-\frac{1}{\tau_2}\mathring{c}_{\mu\nu}\, ,
\label{eq:cevol}
\nd
where $\overline{c}$ and $\mathring{c}$ are defined as the isotropic and traceless parts of the tensor variable $c_{\mu\nu}$ defined as
\bg
\overline{c}_{\mu\nu}=\frac{1}{3}(c^{\lambda}_\lambda-1)(\eta_{\mu\nu}+u_\mu u_\nu)\, , \\
c_{\mu\nu}+u_\mu u_\nu=\mathring{c}_{\mu\nu}+\overline{c}_{\mu\nu}\, .
\nd

In the limit that the relaxation times ($\tau_0, \tau_2$) are very small the evolution equation yields
\bg
c^{ij}=\tau_2( \partial_i u^j+\partial_j u^i-\frac{2}{3}\delta^{ij}\partial_k u^k)+\frac{2}{3}\tau_0\delta^{ij}\partial_k u^k\, .
\label{cijequ}
\nd
Substituting the above equation into $T^{ij}_{LRF}$ and comparing the result to the Navier-Stokes equation (\ref{eq:pimunu}) the bulk and shear viscosities can be identified as
\bg
\eta=\tau_2 p \alpha \nonumber\, , \\
\zeta=\frac{2}{3}\tau_0 p \alpha\, .
\nd
In the model proposed by Ottinger \cite{Ottinger}  
the quantity $\alpha$ is related
to the equation of state,  but in the
linearized version it is simply treated as a constant parameter related
to the relaxation time. 
We fix $\alpha=0.7$ in all calculations, which
then fixes the relaxation times ($\tau_2, \tau_0$) as a function of
$\eta$ and $\zeta$.  The effects of varying $\alpha$ is
shown in appendix \ref{smalltime}.  

It is natural to ask what is the effect of the relaxation time on the
theory.  In some sense this was already answered by Lindblom
\cite{Lindblom}.  
He showed that
the physical fluid must relax to a state that is indistinguishable from the
Navier-Stokes form.  
Therefore we expect the
physical velocity gradients to agree with those given by the auxiliary
tensor variable $c^{\mu\nu}$ as in \Eq{cijequ} .  This is shown in Appendix \ref{sec:grad} for various values of $\eta/s$.  
We expect higher order
gradient terms to be necessary when there are large deviations between
any observable computed using the physical fields or the auxiliary
field $c^{\mu\nu}$.  This will be used as a gauge in order to find the
limit of applicability of hydrodynamics

\subsubsection{1+1 Dimensions}

We now outline the equations of motion for the stress-energy tensor and
the generalized mechanical force tensor assuming a boost-invariant
expansion as well as azimuthal symmetry with arbitrary transverse
expansion.  It is easiest to work in polar coordinates
$(\tau,r,\phi,\eta)$ and since there is no dependence on $\phi$ or
$\eta$ the four-velocity can be expressed as $u^\mu=(\gamma,\gamma
v_r,0,0)$ where $\gamma=\frac{1}{\sqrt{1-v_r^2}}$.  In this coordinate
system the metric tensor is given by,
$g^{\mu\nu}=\text{diag}(-1,1,1/r^2,1/\tau^2)$

The first two equations of motion are given by the conservation of energy and momentum, $d_\mu T^{\mu\nu}=0$ for $\nu=\tau$ and $\nu=r$.  (Due to boost invariance and azimuthal symmetry the $\nu=\eta$ and $\nu=\phi$ equations are trivial.)
\bg
\partial_\tau T^{00}+\partial_r T^{01}=\frac{-1}{\tau}( T^{00}+\Tilde{P}^{33})-\frac{1}{r} T^{01}  \\
\partial_\tau T^{01}+\partial_r T^{11}=\frac{-1}{\tau} T^{01} -\frac{1}{r}( T^{11}-\Tilde{P}^{22}) 
\label{eq:eqom1dT}
\nd
where $\Tilde{P}^{22}=r^2 P^{22}$ and $\Tilde{P}^{33}=\tau^2 P^{33}$.  The evolution equations for the generic mechanical force tensor $c^{\mu\nu}$ are: 
\bg
\partial_\tau c^{11}+v\partial_r c^{11}-\frac{2}{\gamma}[(1-c^{11})\partial_r u^1+c^{01}\partial_r u^0]=\frac{-1}{\gamma\tau_0}\overline{c}^{11}-\frac{1}{\gamma\tau_2}\mathring{c}^{11}   \\
\partial_\tau \Tilde{c}^{22}+v\partial_r \Tilde{c}^{22}+\frac{2v}{r}(\Tilde{c}^{22}-c^{11})+\frac{2}{r}c^{10}=\frac{-1}{\gamma\tau_0}\overline{\Tilde{c}}^{22}-\frac{1}{\gamma\tau_2}\mathring{\Tilde{c}}^{22}   \\
\partial_\tau \Tilde{c}^{33}+v\partial_r \Tilde{c}^{33}+\frac{2}{\tau}(\Tilde{c}^{33}+c^{00})-\frac{2v}{\tau}c^{10}=\frac{-1}{\gamma\tau_0}\overline{\Tilde{c}}^{33}-\frac{1}{\gamma\tau_2}\mathring{\Tilde{c}}^{33} 
\label{eq:eqom1dc}
\nd
where $\Tilde{c}^{22}=r^2 c^{22}$ and $\Tilde{c}^{33}=\tau^2 c^{33}$.

\subsubsection{1+2 Dimensions}

We now consider the 1+2 dimensional case without azimuthal symmetry but
still having longitudinal boost invariance and use a coordinate system
whereby the coordinates transverse to the beam axis are cartesian,
$(\tau,x,y,\eta)$.  Since there is no dependence on $\eta$ the
four-velocity can be expressed as $u^\mu=\gamma(1, v_x,v_y,0)$ where
$\gamma=\frac{1}{\sqrt{1-v_x^2-v_y^2}}$.  In this coordinate system the
metric tensor is given by,
$
g^{\mu\nu}=\text{diag}(-1,1,1,1/\tau^2)
$

In this coordinate system the first three equations of motion are given by the $\nu=\tau$, $x$, and $y$ components of the conservation law $d_\mu T^{\mu\nu}=0$:
\bg
\label{eq:2dstress}
\partial_\tau T^{00}+\partial_x T^{01}+\partial_y T^{02}=\frac{-1}{\tau}( T^{00}+\tau^2 P^{33} )  \\
\partial_\tau T^{10}+\partial_x T^{11}+\partial_y T^{12}=\frac{-1}{\tau}T^{10}\\
\partial_\tau T^{20}+\partial_x T^{21}+\partial_y T^{22}=\frac{-1}{\tau} T^{20}
\nd
The evolution equations for the generalized mechanical force tensor are:
\bg
(\partial_\tau+v_x\partial_x +v_y\partial_y)c^{11}+2[(c^{11}-1)\partial_x v_x+c^{12}\partial_x v_y]=\frac{-1}{\gamma\tau_0}\overline{c}^{11}-\frac{1}{\gamma\tau_2}\mathring{c}^{11}   \\
(\partial_\tau+v_x\partial_x +v_y\partial_y)c^{22}+2[(c^{22}-1)\partial_y v_y+c^{21}\partial_y v_x]=\frac{-1}{\gamma\tau_0}\overline{c}^{22}-\frac{1}{\gamma\tau_2}\mathring{c}^{22}   \\
(\partial_\tau+v_x\partial_x +v_y\partial_y)\Tilde{c}^{33}+\frac{2}{\tau}(\Tilde{c}^{33}-1)=\frac{-1}{\gamma\tau_0}\overline{\Tilde{c}}^{33}-\frac{1}{\gamma\tau_2}\mathring{\Tilde{c}}^{33} \\
(\partial_\tau+v_x\partial_x +v_y\partial_y)c^{12}+c^{12}(\partial_x v_x + \partial_y v_y)+(c^{22}-1)\partial_x v_y+(c^{11}-1)\partial_y v_x\nonumber\\=\frac{-1}{\gamma\tau_0}\overline{c}^{12}-\frac{1}{\gamma\tau_2}\mathring{c}^{12}  
\label{eq:2dcs}
\nd

\subsubsection{Initial Conditions}
\label{sec:IC}

The hydrodynamic simulation is a $2+1$ boost invariant
hydrodynamic model with an ideal gas equation
of state $p = \frac{1}{3}\,\epsilon$. 
The temperature is
related to the energy density with the $N_f=3$ ideal
QGP equation of state.  
We have chosen this
extreme equation of state because the resulting radial
and elliptic flow are too large relative to data on
light hadron production. Thus,
this equation of state will estimate the largest elliptic
flow possible for a given shear viscosity.  We note that for any non-central collision we have choosen a default impact parameter of b=6.5 fm.   

Aside from the equation of state, the hydrodynamic model is based upon
reference \cite{Teaney:2001av}.
At an initial time $\tau_0$,
the entropy is distributed in the transverse plane
according to the distribution of participants
for a Au-Au collision.
Then one parameter, $C_s$,
is  adjusted to set the initial temperature and total particle yield.
Specifically  the initial entropy density in the transverse plane is
\st
    s(x, y,\tau_0) = \frac{C_{s}}{\tau_0}\,\frac{dN_{p}}{dx\,dy},
\stp
where $\frac{dN_p}{dx\,dy}$ is the number of participants per
unit area.
The value  $C_{s}=15$ closely corresponds to the
results of full hydrodynamic simulations \cite{Teaney:2001av,Kolb:2000fh,Huovinen:2001cy} and  corresponds to a maximum initial
temperature of $T_{0} = 420\,\mbox{MeV}$ at impact
parameter $b=0$. 
With the entropy density specified the energy density can be 
determined. This requires inverting the equation of 
state.

In a viscous formulation we must also specify the viscous 
fields, $i.e.$ the  $c^{\mu\nu}$ in the second order setup.
Following the general philosophy outlined in \Sect{viscoushydro} we
will choose the $c^{\mu\nu}$ such that the stress tensor
deviations are
\st
 \pi_{\mu\nu} = -\eta\llangle \nabla_{\mu} u_{\nu} \rrangle  
\qquad  \Pi  = -\zeta \nabla_{\mu} u^{\mu} = 0 
\stp
Since at time $\tau_o$ the transverse flow velocity and the longitudinal
flow velocity is Bjorken this means that
at mid rapidity  
\st
\pi_{xx} = \pi_{yy} =  -\frac{1}{2} \,\pi_{zz}= \frac{2}{3}\, \eta\, \partial_z u^{z}  \qquad \Pi = 0
\stp

To achieve this condition we first rewrite the flow equations 
for small $c_{\mu\nu}$ and vanishing transverse flow. The 
$c_{ij}$ equations become
\bg
   \partial_\tau c^{11}  &=&  
- \frac{\bar{c}^{11}}{\tau_0}
- \frac{\mathring{c}^{11}}{\tau_2}\, , \\
   \partial_\tau c^{22}  &=&  
- \frac{\bar{c}^{22}}{\tau_0}
- \frac{\mathring{c}^{22}}{\tau_2}\, , \\
   \partial_\tau c^{33}  - \frac{2}{\tau} &=&  
- \frac{\bar{c}^{33}}{\tau_0}
- \frac{\mathring{c}^{33}}{\tau_2}\, . 
\nd
In writing this we have used the fact that for small velocity
$c^{00} \approx  -u^{0} u^{0}$ .
Then looking for the quasi stationary state we set the time derivatives
to zero, and  use the relations $\bar{c}^{ij} = \frac{1}{3} c^l_l\,\delta^{ij}$ and 
$c^{ij} = \mathring{c}^{ij} + \bar{c}^{ij}$ to find that 
\bg
  c^{11}&=& \frac{2}{3} \frac{\tau_0}{\tau} - \frac{2}{3} \frac{\tau_2}{\tau}\, , \\
  c^{22}&=& \frac{2}{3} \frac{\tau_0}{\tau} - \frac{2}{3} \frac{\tau_2}{\tau}\, ,\\
  c^{33}&=& \frac{2}{3} \frac{\tau_0}{\tau} + \frac{4}{3} \frac{\tau_2}{\tau}\, .
\nd

\section{Hydrodynamic Results}
\label{sec:HydroResults}
The equations outlined in the previous two sections were integrated
numerically using the initial conditions described above.  The
algorithm \cite{Pareschi} and a discussion of the numerics can be found in appendix
\ref{sec:Alg}.  In this section we now show the results of the
simulation.  Before showing the results of the  2+1 dimensional
simulation we outline some of the main physics points using results
from the 1+1 dimensional case.

\begin{figure}[hbtp]
  \vspace{9pt}
  \centerline{\hbox{ \hspace{0.0in}
\includegraphics[scale=0.65]{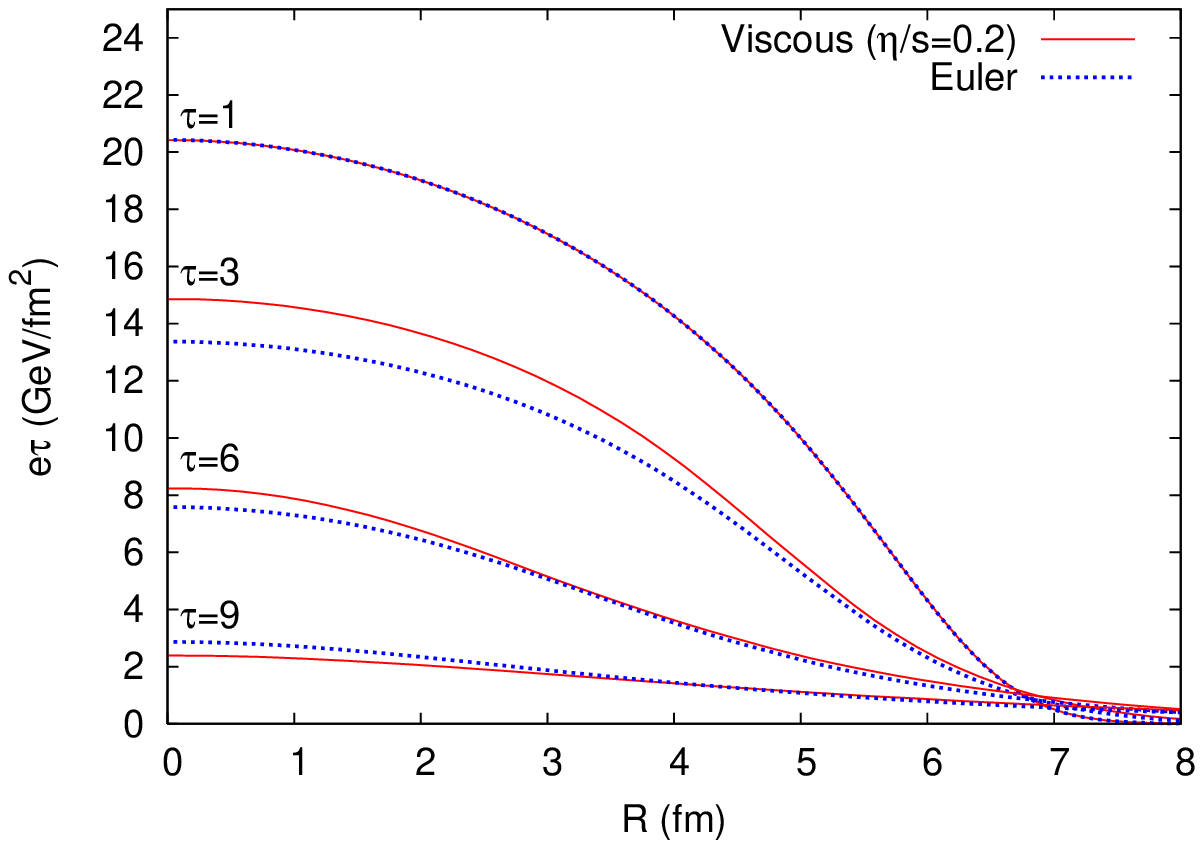}
    \hspace{0.0in}
\includegraphics[scale=0.65]{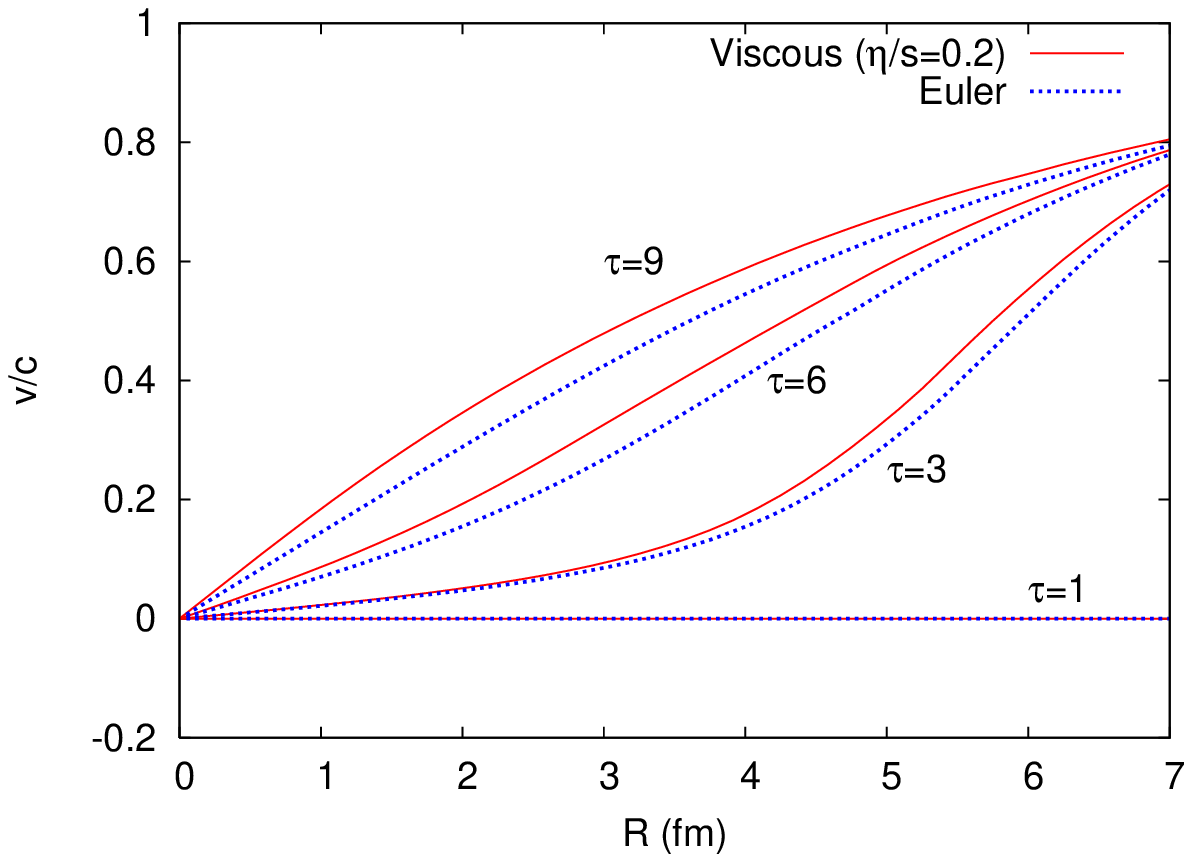}
    }
  }
\caption{(Color online) Plot of the energy density per unit rapidity (left) and of the
transverse velocity (right) at times of $\tau=1,3,6,9$ fm/c, for
$\eta/s=0.2$ (solid red line) and for ideal hydrodynamics (dotted blue line).}
\label{fig:v1d}
\end{figure}

Fig.~\ref{fig:v1d} shows the energy density per unit rapidity (left)
and the transverse velocity (right) at various times for both ideal
hydrodynamics and for finite viscosity ($\eta/s=0.2$).  
The effect of viscosity is twofold. The longitudinal 
pressure is initially reduced and the viscous case does less longitudinal $pdV$ work as in the simple Bjorken expansion \cite{Danielewicz:1984ww}.  This means that at early times the energy per rapidity decreases
more slowly in the viscous case.
The reduction of longitudinal pressure is 
accompanied by a larger transverse pressure.
This causes the transverse velocity
to grow more rapidly.  The larger transverse velocity causes the energy
density to deplete faster at later times in the viscous case.  The net result is that a
finite viscosity (even as large as $\eta/s=0.2$) does not integrate to
give major deviations from the ideal equations of motion. A preliminary account of this effect was given long ago \cite{DT1}.

\begin{figure}[hbtp]
  \vspace{9pt}
  \centerline{\hbox{ \hspace{0.0in} 
\includegraphics[scale=1.2]{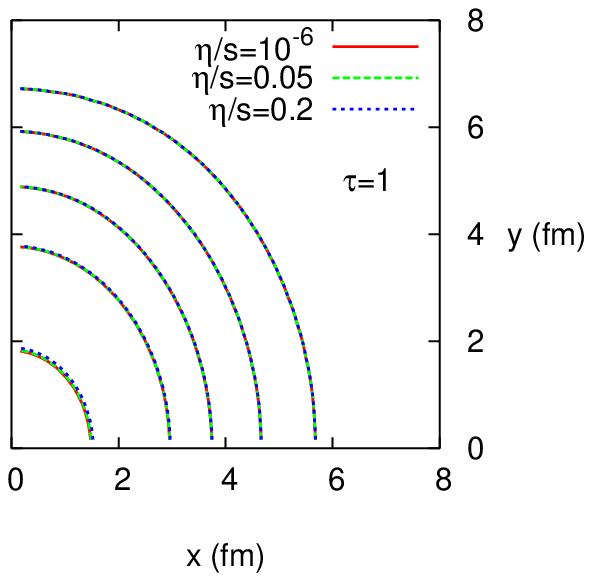}
    \hspace{0.1in}
\includegraphics[scale=1.2]{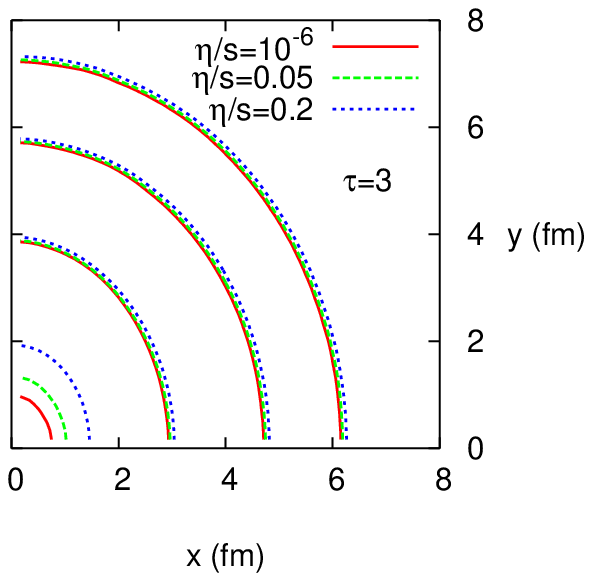}
    }
  }
  \vspace{9pt}
  \centerline{\hbox{ \hspace{0.0in}
\includegraphics[scale=1.2]{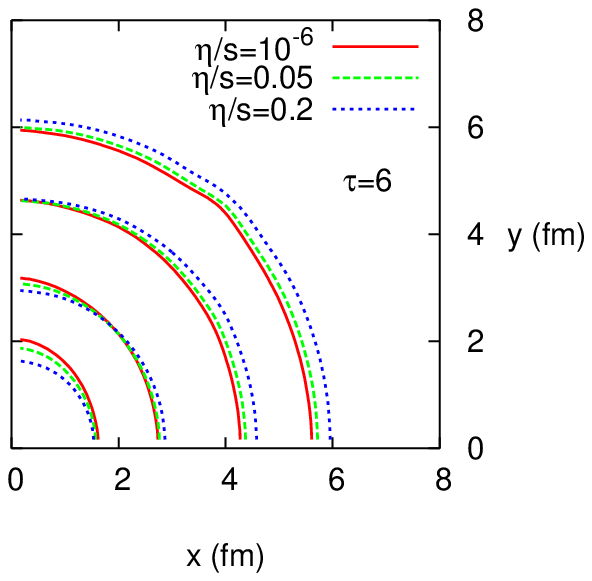}
    \hspace{0.1in}
\includegraphics[scale=1.2]{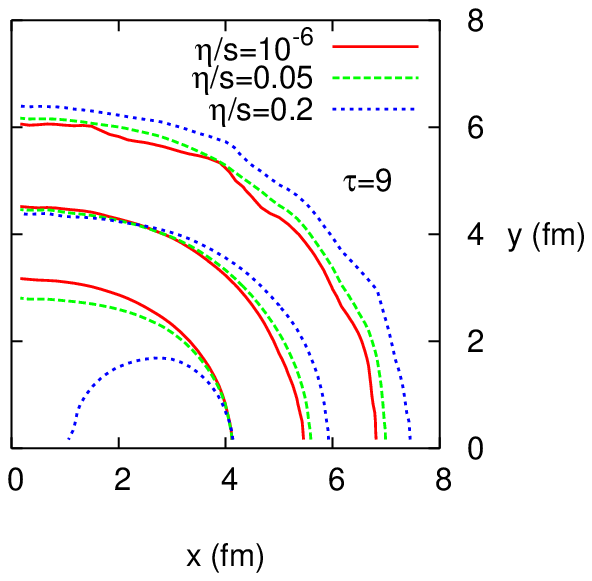}
    }
  }
  \vspace{9pt}
  \caption{(Color online) Contour plot of energy density per unit
rapidity in the transverse plane.  The contour values working outward
are for $\tau=1$ fm/c: 15, 10, 5, 1, 0.1, for $\tau=3$ fm/c: 10, 5, 1, 0.1, for $\tau=6$ fm/c: 3,
2, 1, 0.5 and for $\tau=9$ fm/c: 0.5, 0.375, 0.25, in units of GeV/fm$^2$.  }
  \label{fig:enedens}
\end{figure}

We now present results of the 2+1 dimensional boost invariant
hydrodynamic model.  Fig.~\ref{fig:enedens} shows contour plots of the
energy density per unit rapidity in the transverse plane at proper
times of $\tau=1, 3, 6, 9$ fm/c.  The initial conditions ($\tau=1$) is taken
from the Glauber model discussed before.

\begin{figure}[hbtp]
  \vspace{9pt}
  \centerline{\hbox{ \hspace{0.0in} 
\includegraphics[scale=1.2]{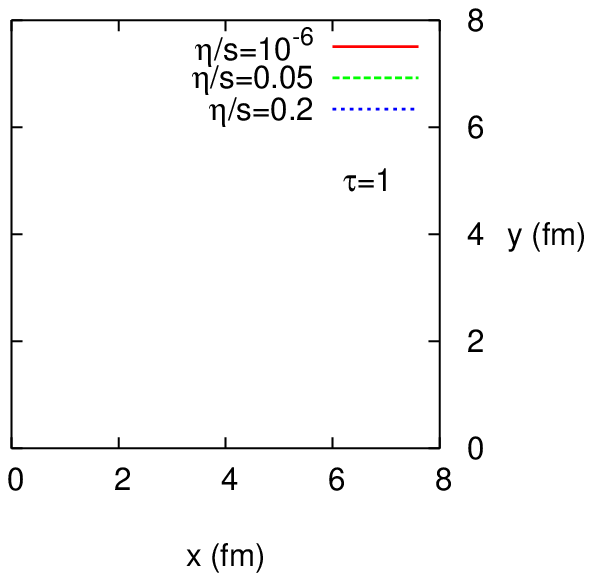}
    \hspace{0.1in}
\includegraphics[scale=1.2]{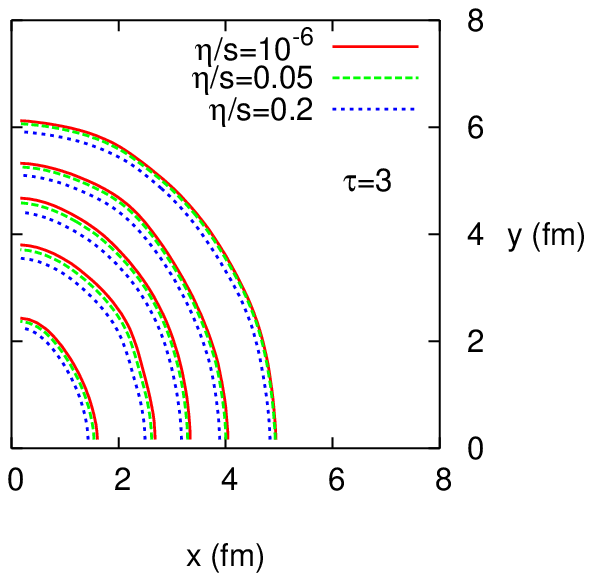}
    }
  }
  \vspace{9pt}
  \centerline{\hbox{ \hspace{0.0in}
\includegraphics[scale=1.2]{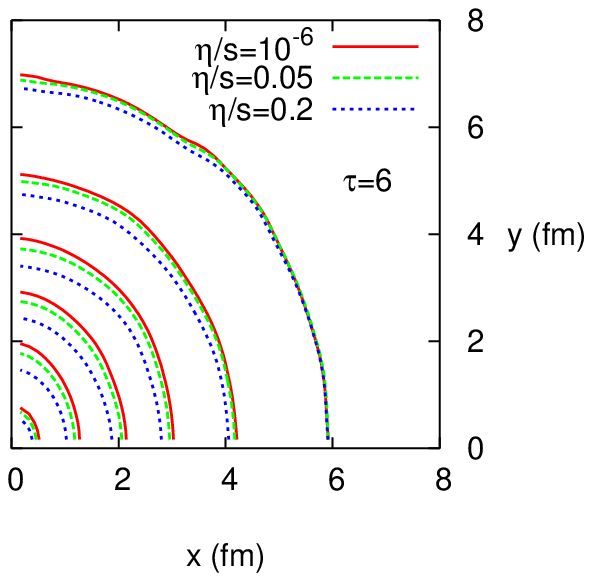}
    \hspace{0.1in}
\includegraphics[scale=1.2]{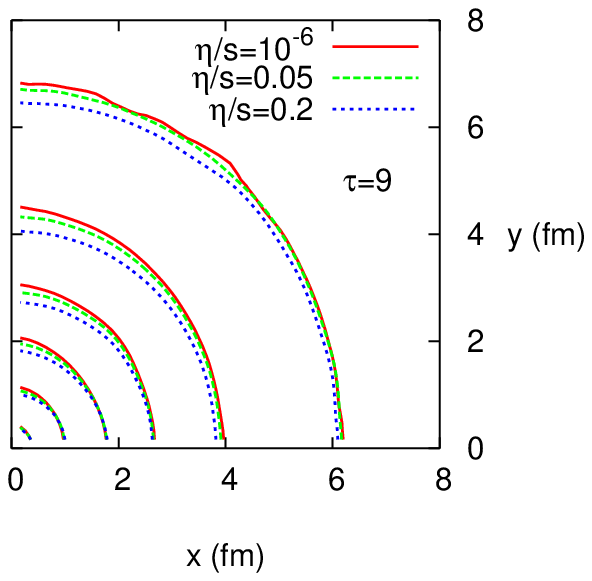}
    }
  }
  \vspace{9pt}
  \caption{(Color online) Contour plot of transverse velocity, $v_\perp=\sqrt{v_x^2+v_y^2}$.  The inner most contour is for $v_\perp=0.1$ and increases in steps of $\Delta v_\perp = 0.15$. }
  \label{fig:vt}
\end{figure}

Fig.~\ref{fig:vt} shows contour plots of the transverse velocity at the
same times of $\tau=$1, 3, 6, 9 fm/c.  At $\tau=1$ the figure is blank since
the velocity in the transverse plane is zero as set by the initial
conditions.  By looking at the contours of constant $v/c$ one can see
that a finite viscosity increases the transverse velocity.

\begin{figure}[hbtp]  
\includegraphics[scale=.7]{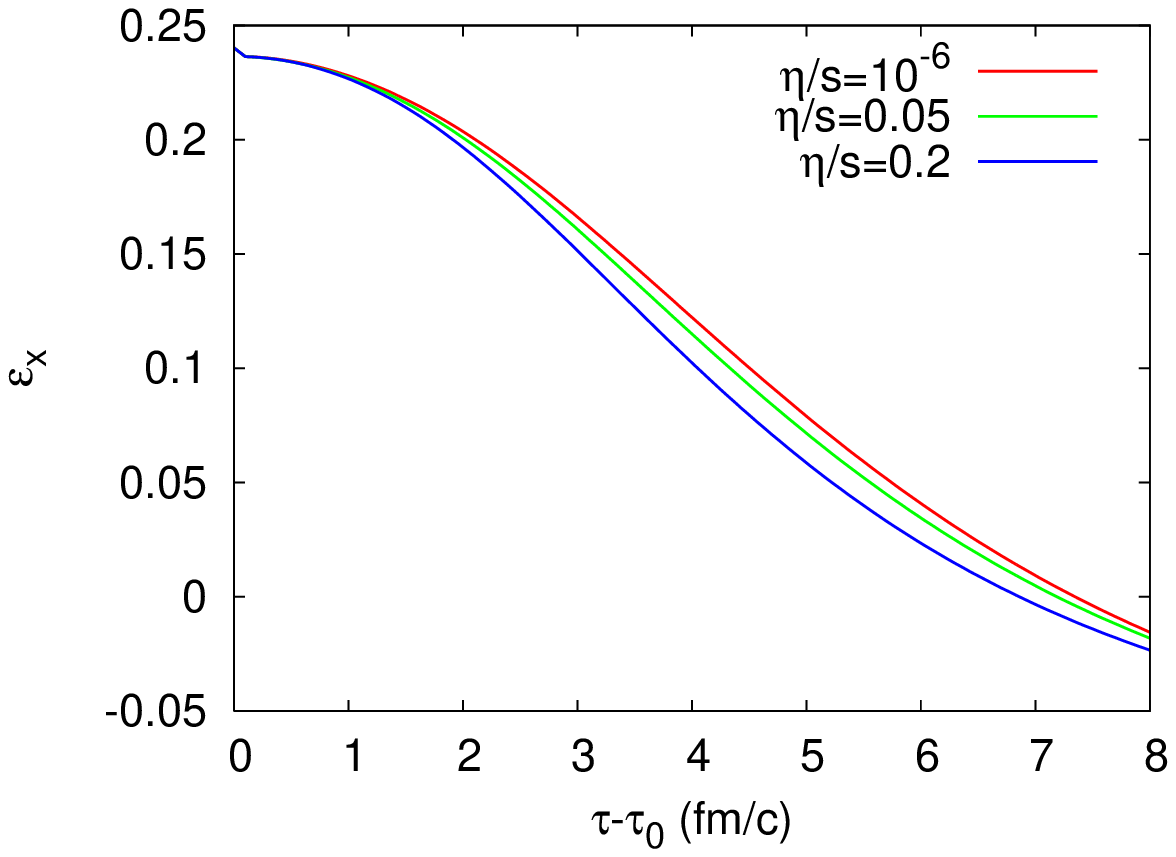}
 \vspace{9pt}
\includegraphics[scale=.7]{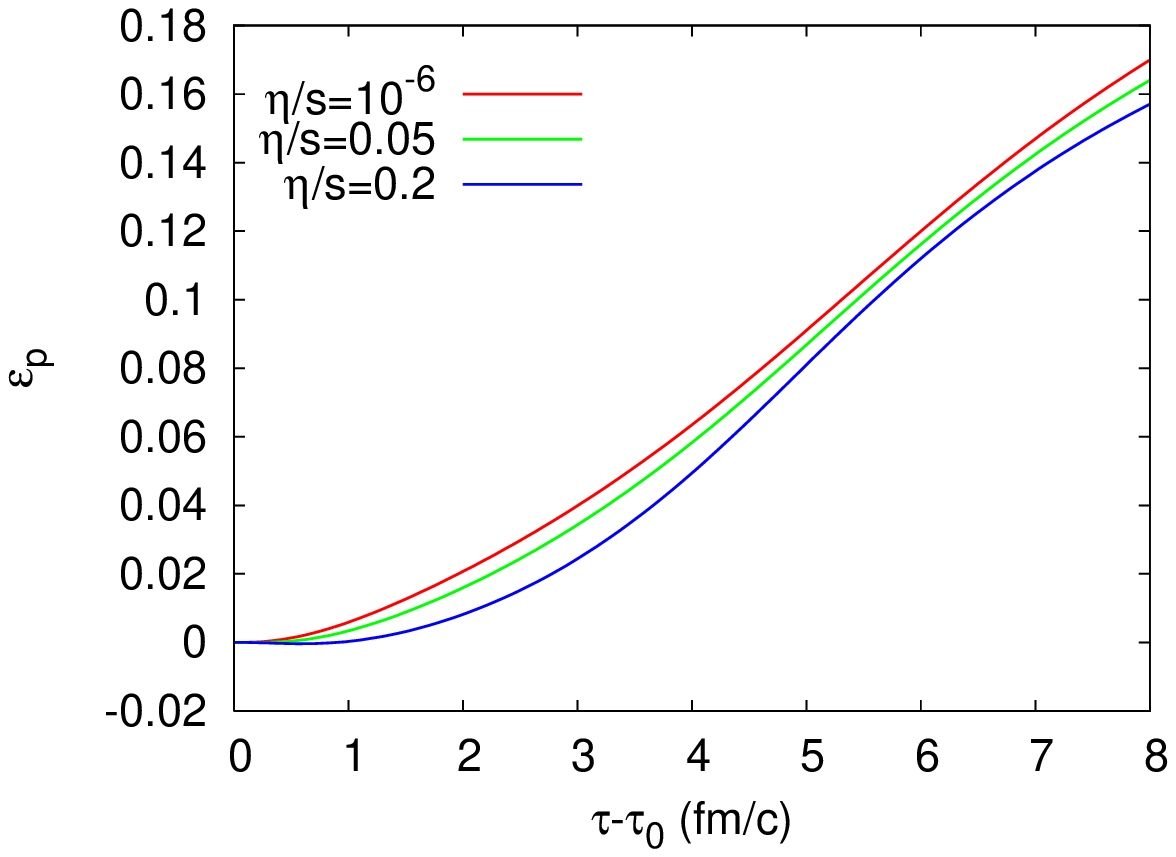}
\vspace{9pt}
\includegraphics[scale=.7]{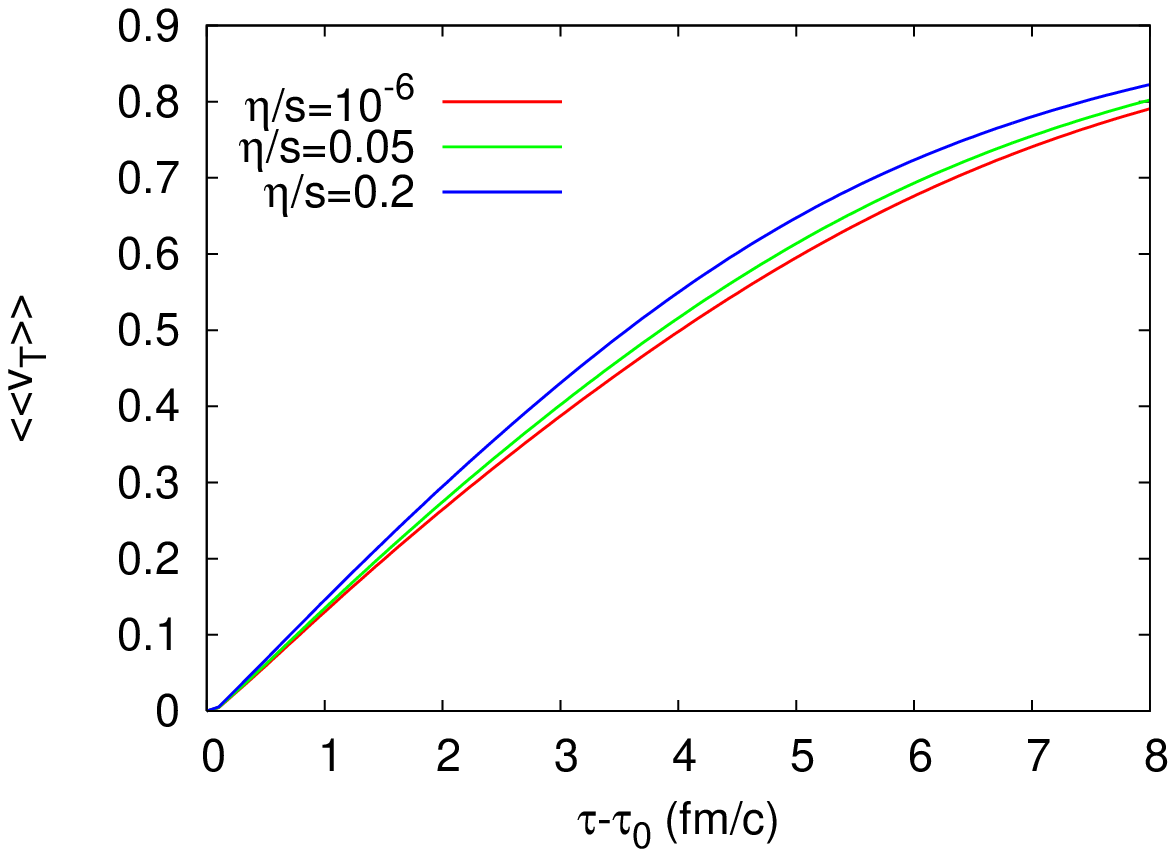}
  \vspace{9pt}
  \caption{(Color online) Time evolution of the spatial ellipticity
$\epsilon_x$, the momentum anisotropy $\epsilon_p$, and the energy
density weighted transverse flow $\langle\langle v_\perp
\rangle\rangle$, see Eq.~\ref{eq:anis}. }
  \label{fig:anis}
\end{figure}

Since we are interested in elliptic flow which originates from the
initial spatial anisotropy of the collision region it is useful to see
how the spatial and momentum anisotropy develop in time.  We therefore
look at the following three quantities \cite{KSH}:
\bg
\epsilon_x=\frac{\langle\langle y^2-x^2 \rangle\rangle}{\langle\langle y^2+x^2 \rangle\rangle} \nonumber\\
\epsilon_p=\frac{\langle\langle T^{xx}-T^{yy} \rangle\rangle}{\langle\langle T^{xx}+T^{yy} \rangle\rangle} \nonumber\\
\langle\langle v_T \rangle\rangle=\frac{\langle\langle \gamma \sqrt{v_x^2+v_y^2} \rangle\rangle}{\langle\langle \gamma \rangle\rangle}
\label{eq:anis}
\nd
where the double angular bracket $\langle\langle \cdots \rangle\rangle$
denote an energy density weighted average.  The spatial ellipticity
($\epsilon_x$) is a measure of the spatial anisotropy as a function of
time.  The spatial anisotropy is what drives the momentum anisotropy
($\epsilon_p$).  This quantity can be thought of as characterizing the
$p_T^2$ weighted integrated elliptic flow \cite{OllitraultSph}. 
 The final quantity $\langle\langle v_T \rangle\rangle$ is the average
radial flow velocity.  All three of these quantities are plotted in
fig.~\ref{fig:anis} for $\eta/s$=0.2, 0.05 and $10^{-6}$.

As already shown in the 1+1 dimensional case the finite viscosity case
does less longitudinal work. The longitudinal pressure is reduced
while the transverse pressure is uniformly increased in the radial
direction, i.e. gives no addition $v_2$ component.
This causes the transverse velocity (as seen in
$\langle\langle v_T \rangle\rangle$ and fig.~\ref{fig:vt}) to grow more
rapidly while $\epsilon_p$ lags behind the ideal case. 
Furthermore, the larger radial symmetric transverse velocity 
causes a faster decrease in
the spatial anisotropy. This further frustrates the build-up 
of the momentum anisotropy $\epsilon_p$.
 We therefore expect to see a decrease in the integrated
$v_2$ as the viscosity is increased.  This is indeed the case as will
be shown.  However, this effect is small compared to the change in
$v_2$ from modifications of the off-equilibrium distribution function.

\section{Freezeout}
\label{sec:fo}

As discussed in the introduction, ideal hydrodynamics is
applicable when $\lambda_{mfp}\ll L$ where $L$ denotes the typical
system size.  When dissipative corrections are included, one must
remember that the Navier Stokes equations are derived assuming that the
relaxation time $\tau_R$ is much 
smaller than the inverse expansion rate, 
$\tau_R \partial_{\mu} u^{\mu}\ll 1$.
Therefore, in the simulations we determine the freezeout surface by monitoring the expansion rate relative to the relaxation time using a generalization of the freezeout criteria first proposed in \cite{Bondorf:1978kz, Heinz:1987ca} and later in \cite{Hung:1997du}.

Specifically, freezeout is signaled when\footnote{
In actual simulations we take 
$(\eta/p)\,\partial_{\mu} u^{\mu} = 0.6$ for most runs (see below).
}
\st
     \frac{\eta}{p} \partial_{\mu} u^{\mu} \sim \frac{1}{2}
\stp
This combination  of parameters can be motivated from the kinetic 
theory estimates \cite{Reif}.  The pressure is $p\sim \epsilon
\llangle v_{\rm th}^2\rrangle  $  with $\llangle v_{\rm th}^2\rrangle $ the 
typical quasi-particle velocity and $\epsilon$ the energy density.
The viscosity is of order $\eta \sim \epsilon \llangle v_{\rm th}^2 \rrangle \tau_{R}$ 
with $\tau_R$ the relaxation time.  Thus the freezeout condition is
simply
\st
    \frac{\eta}{p} \partial_{\mu} u^{\mu} \sim \tau_R \partial_{\mu} u^{\mu} \sim \frac{1}{2} 
\stp
In the model we are considering $\eta/p=\alpha \tau_{2}$ with 
$\alpha =0.7$ as described in \Sect{GenericSect}.

The value of $\frac{1}{2}$ can be considered as a parameter chosen to
be smaller than one.  The point is that as the above quantity becomes
large the Navier Stokes approximation is no longer applicable and the
simulation should freezeout.  At this point one would need to include
further higher order corrections in the gradients or switch to a
kinetic approach.


It is also convenient to have a definition for an analogous freezeout
surface in the case of ideal hydrodynamics.  One can think of keeping
the freezeout surface fixed as $\eta/s$ is taken to zero.  Dividing the
freezeout criterion by $\eta/s$ and using $s=(\epsilon+p)/T\sim 4p/T$
we define:
\st
\chi=\frac{4}{T}\partial_\mu u^\mu
\stp
which involves only quantities in the ideal simulation.
This is  a separate freezeout parameter independent of the viscosity. 

\begin{figure}[hbtp]
  \vspace{9pt}
  \centerline{\hbox{ \hspace{0.0in} 
\includegraphics[height=75mm]{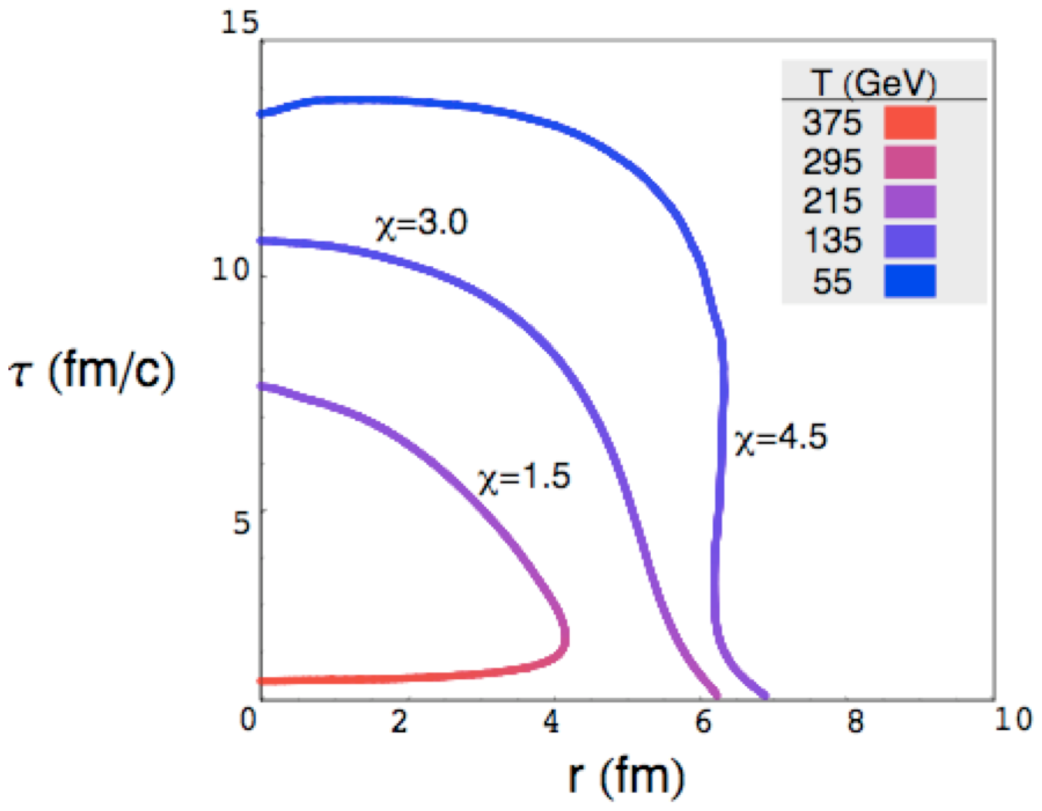}
    \hspace{0.0in}
\includegraphics[height=75mm]{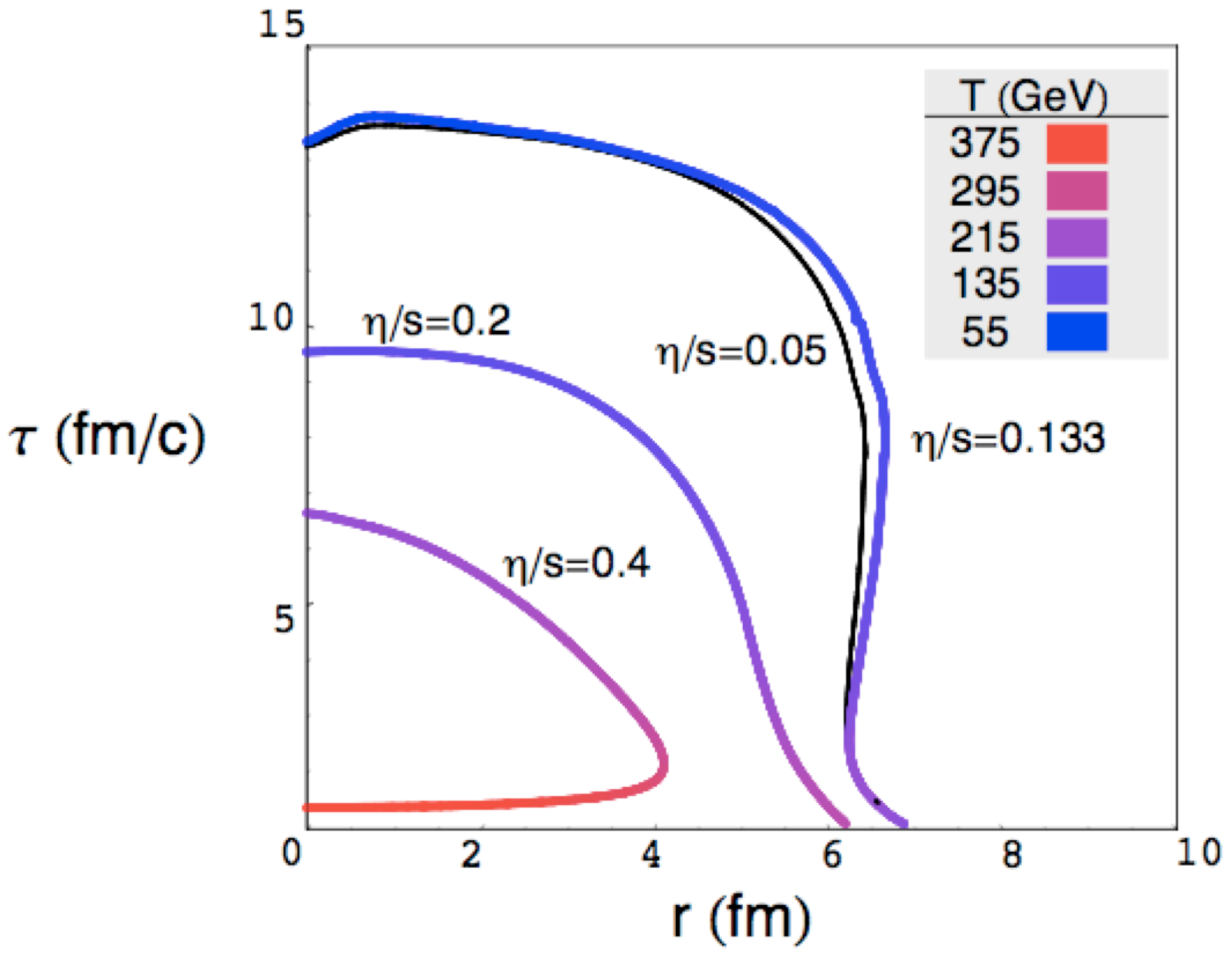}
    }
  }
\caption{(Color online) Contour plot of various freezeout surfaces for central Au-Au
collisions.  Left: Surfaces from ideal hydrodynamics where the freezeout
condition is set by the parameter $\chi$=1.5, 3 and 4.5.  Right:
Corresponding viscous solution where $\eta/s$ was fixed by the
condition $\frac{\eta}{p}\partial_\mu u^\mu=0.6$.  The thin solid black
curve shows the contour set by $\frac{\eta}{p}\partial_\mu u^\mu=0.225$
for comparison.  }
\label{fig:fosurf}
\end{figure}

We show in fig.~\ref{fig:fosurf} contour plots of the freezeout
surface for fixed $\chi$ from both ideal (left plot) and viscous
hydrodynamics (right plot).  For fixed $\chi$ the freezeout surfaces
remain approximately the same in both cases.  The freezeout surface from now on
will be specified by $\chi$ in order to facilitate a comparison between
the ideal and viscous cases when comparing spectra.

We have typically chosen $\chi$ and $\eta/s$ in order that
$\frac{\eta}{p}\partial_\mu u^\mu=0.6$. 
Thus in Table~\ref{tabFO} for $\eta/s=0.2$ we have $\chi=3.0$
and $\frac{\eta}{p}\partial_\mu u^\mu=0.6$. 
However, for $\eta/s=0.05$ the
freezeout parameter is $\chi=12$ giving an unphysically large
surface. 
This would normally not be the case in a more
realistic model with a phase transition present, since in the hadronic
phase the viscosity goes like $\eta\sim\frac{T}{\sigma_0}$.  The change
in scaling with temperature would cause the system to freezeout soon
after hadronization.  We plan on quantifying this statement in a future
work.  We therefore use $(\eta /p)\partial_\mu u^\mu=0.225$ when
$\eta/s=0.05$ giving $\chi=4.5$.  The thin solid curve in the right
plot of fig.~\ref{fig:fosurf} shows this particular freezeout contour.     
In table~\ref{tabFO} we summarize the freezeout parameters used
throughout this work.  For a given $\eta/s$ the most physical
choice of freezeout parameter $\chi$ is selected such that $(\eta /p)\partial_\mu u^\mu\approx0.6$.               
However, if the viscosity becomes so
small that the volume becomes unphysically large (such as for $\eta/s=0.05$) we
set $\chi=4.5$ as a maximum.  These three physically motivated 
parameter sets are given in bold  in the table.

We should stress that the freezeout surface taken in this work is
different from the typical constant temperature surface
used in many hydrodynamic simulations.  From fig.~\ref{fig:fosurf}, one
can see from the temperature map that the surface is not an
isotherm and actually spans a very wide range of temperatures.  
The freezeout surface is understood by examining the expansion rate in Bjorken geometry
\st
   \partial_{\mu} u^{\mu} = \partial_{\tau} u^{\tau} + \frac{u^{\tau}}{\tau} + \partial_{x} u^{x} + \partial_y u^{y}  \, .
\label{eq:grad}
\stp
The
resulting surface is due to a competition 
between the first two terms
in \ref{eq:grad} at early times and the last two terms at later times 


\begin{table}[hbtp]
\begin{tabular}{c|c|c}
\hline
$\eta/s$ & $\frac{\eta}{p}\partial_\mu u^\mu$ & $\chi$\\
\hline
0.05 & 0.6 & 12.0 \\ 
{\bf 0.05} & {\bf 0.225} & {\bf 4.5} \\
0.05 & 0.15 & 3.0 \\
\hline
0.2 & 0.9 & 4.5 \\
{\bf 0.2} & {\bf 0.6} & {\bf 3.0} \\
\hline
{\bf 0.133} & {\bf 0.6} & {\bf 4.5} \\ 
\hline
\end{tabular}
\caption{Freezeout parameters used throughout this work.  For a given $\eta/s$ the most physical
choice of freezeout parameter $\chi$ is selected such that $(\eta /p)\partial_\mu u^\mu\approx0.6$. 
However, if the viscosity becomes so small (such as for $\eta/s=0.05$) that the volume becomes unphysically large (see text for discussion) we
set $\chi=4.5$ as a maximum.  These three physically motivated
parameter sets are in bold.}
\label{tabFO}
\end{table}

\section{Spectra}

\subsection{Anisotropy}

Before computing the differential spectrum we will compute the 
momentum anisotropy as a function of time. The momentum 
anisotropy $A_2$ (which differs from $v_2$ by the placement 
of averages) is defined as
\bg
A_2=\frac{\langle p_x^2\rangle-\langle p_y^2\rangle}{\langle p_x^2\rangle+\langle p_y^2\rangle}= \frac{S_{11}-S_{22}}{S_{11}+S_{22}}\, ,
\label{eq:A2} 
\nd
where $S^{ij}$ is the sphericity tensor and can be related to the 
hydrodynamics fields (i.e. $u^{\mu}$ ,$\pi^{\mu\nu}$, $\Pi$) 
and moments of the ideal particle distribution function. 
The explicit form is given in appendix \ref{sec:sph} and generalizes
an appendix of Ollitrault \cite{OllitraultSph} to the viscous case.
From a theoretical perspective, $A_2$ is preferred because it is
almost independent of the details of the particle content of the theory
\cite{OllitraultSph}.

We plot $A_2$ in the following manner.  At a given proper time we integrate over the surface of constant $\chi$, 
which has developed by time $\tau$.  The remaining part of the surface is fixed by integrating
over the matter which has not frozen out ($\chi < \chi_{f.o.}$) at fixed
proper time.  This can be thought of as a freezeout surface with a flat top at time $\tau$. As time
moves forward eventually all of the matter is frozen out over a surface set by constant $\chi$ yielding a constant $A_2$.

\begin{figure}[hbtp]
  \vspace{9pt}
  \centerline{\hbox{ \hspace{0.0in} 
\includegraphics[scale=0.7]{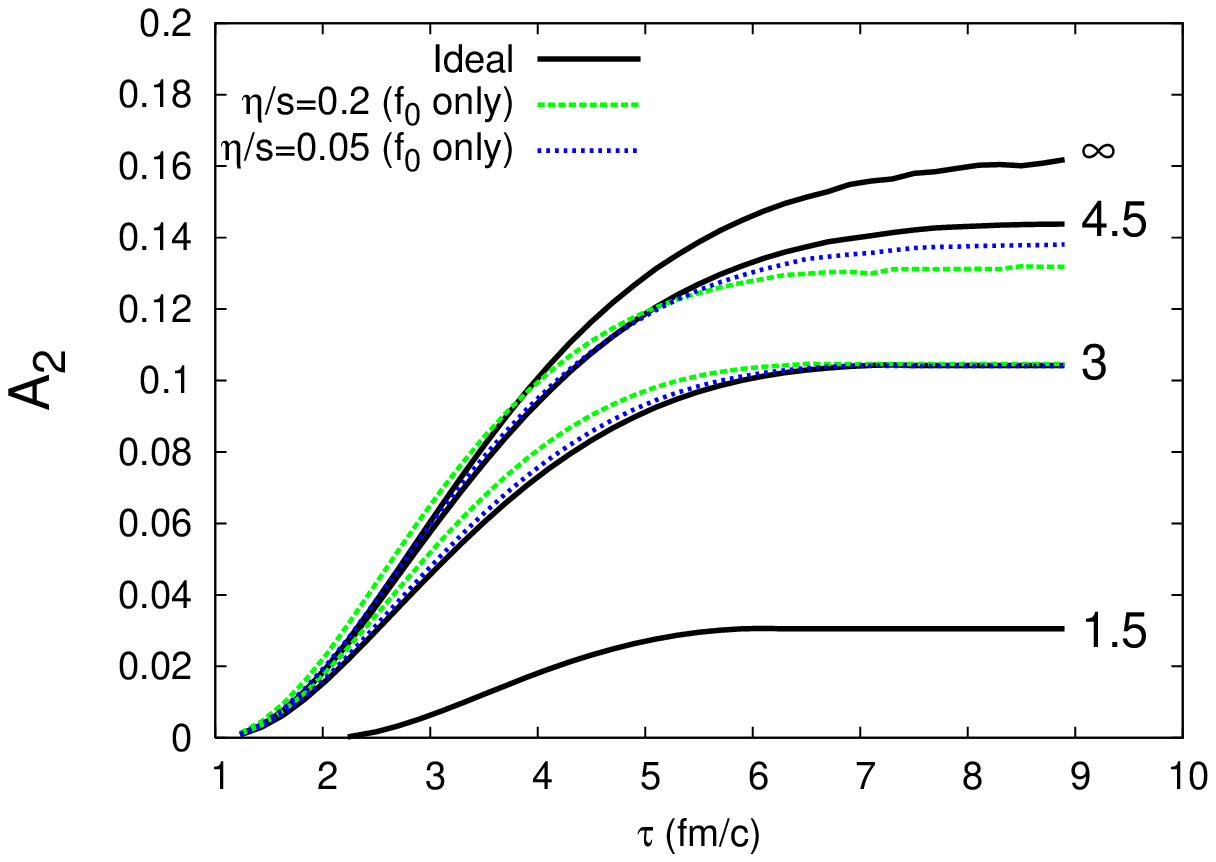}
    \hspace{0.0in}
\includegraphics[scale=0.7]{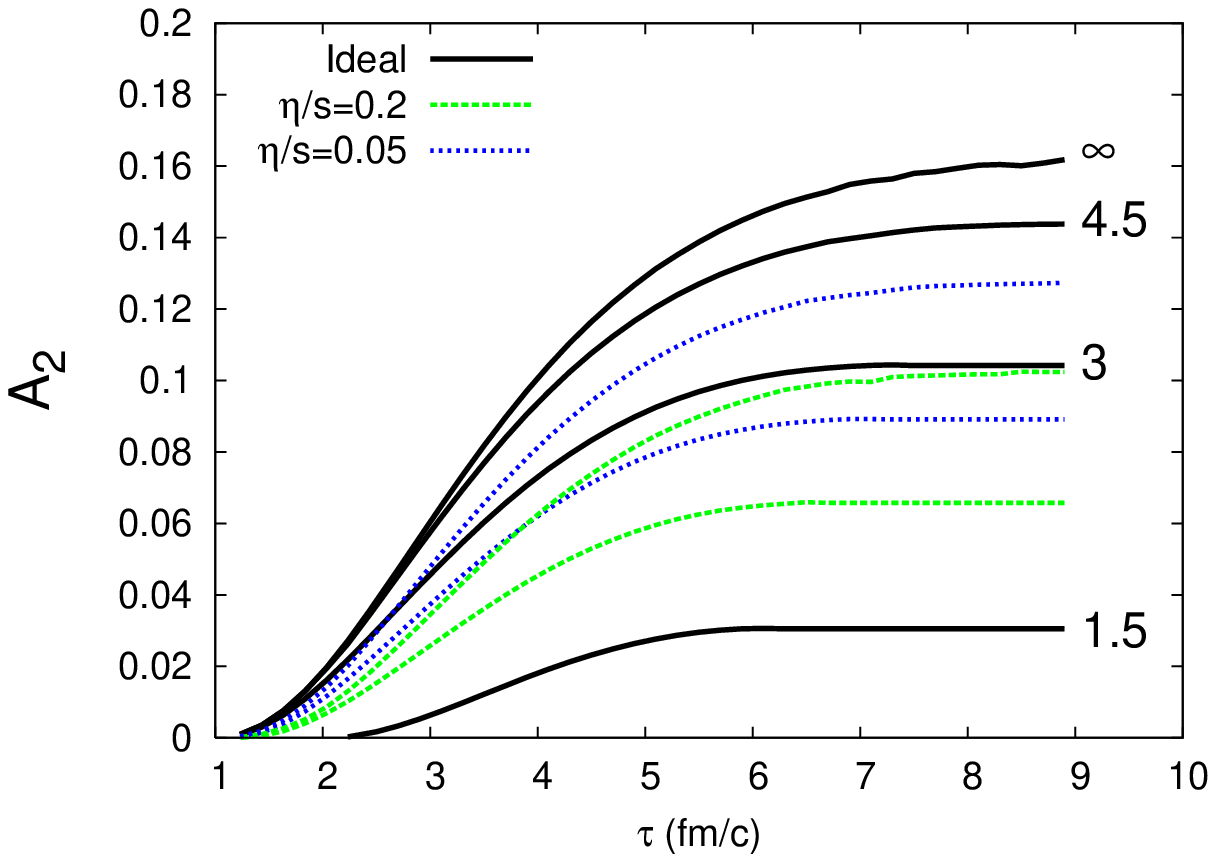}
    }
  }
\caption{(Color online) $A_2$ (defined in Eq.~\ref{eq:A2}) as a function of $\tau$.  The solid black
lines show the ideal result for $\chi=$1.5, 3.0, 4.5 and $\infty$.
Also shown in the right and left figures respectively are the viscous results with 
and without including the viscous correction to the distribution
function, for $\chi=3.0$ and $4.5$ and $\eta/s=0.2$ (dashed green curve) and for $\eta/s$=0.05
(dotted blue curve).} 
\label{fig:A2}
\end{figure}

Figure~\ref{fig:A2} shows $A_2$ for four different freezeout surfaces.
The figure on the left shows the results using only the ideal
contribution to the sphericity (regardless of if viscosity is present).
This will be analogous to using only the ideal particle distribution
function when generating the spectrum.  
First look at the solid black curves which
are generated using ideal hydrodynamics and a specified $\chi$.  
For a larger value of $\chi$
a larger space-time region is evolved by 
hydrodynamics producing a larger elliptic flow or $A_2$. 
The
true ideal case  where hydrodynamics is universally applicable  
is given by $\chi=\infty$.  We see that for $\chi=4.5$ most of
the elliptic flow is reproduced.

In order to assess the role of viscosity we first look at the figure on
the left.  The dashed curves show $A_2$ 
for $\eta/s=0.05$ and $\eta/s=0.2$ without including viscous
corrections to the distribution function. (For clarity, these curves are shown only for $\chi=3.0$ and $\chi=4.5$.)  
Without the corrections to the distribution function the viscous 
corrections to $A_2$ are modest.
The right figure shows the analogous
plot, this time including the viscous corrections to the distribution
function.  The corrections are much larger and we therefore expect the
viscosity to decrease the integrated elliptic flow.    

\subsection{Spectra}

The thermal $p_T$ and differential $v_2$ spectra of particles are
generated using the Cooper-Frye formula \cite{CF} given by
\st
E\frac{d^3N}{d^3p}=\frac{g}{2\pi^3}\int_\sigma f(p_\mu u^\mu,T) p^\mu d\sigma_\mu\, .
\label{eq:CF}
\stp
The thermal distribution function used in the Cooper-Frye formula above also needs to include corrections due to finite viscosity.  We therefore write $f = f_o + \delta f$ where $f_o$ is the ideal particle distribution and $\delta f$ is the viscous correction which has been derived in appendix~\ref{sec:vc} and is given by
\st
\delta f = \frac{1}{2(e + p) T^2}\, f_o(1 + f_o) \, 
p^{\mu} p^{\nu} \left[\pi_{\mu\nu} + \frac{2}{5} \Pi \Delta_{\mu\nu}  \right].
\label{eq:df}
\stp
For boltzmann statistics $f_o(1+f_o)$ is replaced by $f_o$.  
The elliptic flow is defined as the weighted average of the yields with $\cos(2\phi)$:
\st
v2(p_T)=
\langle \cos(2\phi) \rangle_{p_T}=\frac{\int_{-\pi}^\pi d\phi \cos(2\phi) \frac{dN}{dy p_T dp_T d\phi}}{\int_{-\pi}^\pi d\phi\frac{dN}{dy p_T dp_T d\phi}}\, ,
\stp
where $\phi$ is the angle between the decaying particle's momentum
(${\bf p}_T$) and the azimuthal angle of the collision region.

\begin{figure}
\includegraphics[scale=0.8]{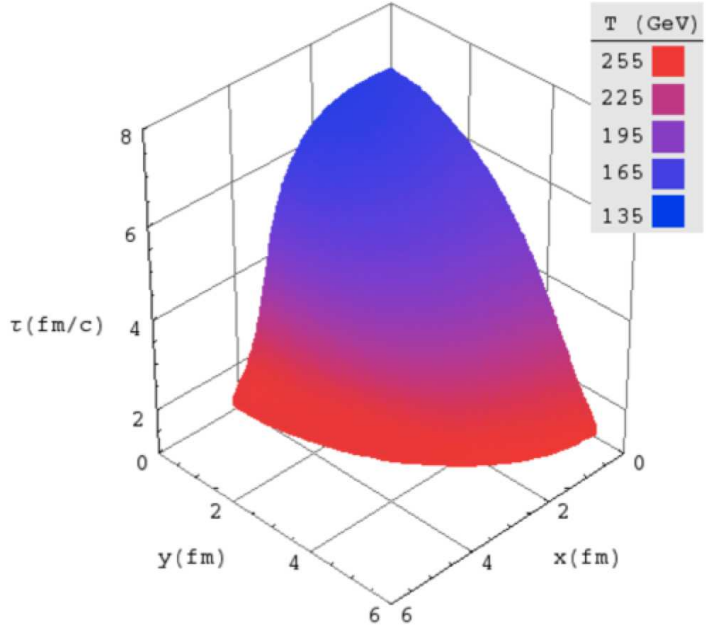}
\caption{(Color online) Freezeout surface for semi-central (b=6.5) Au-Au collisions
for $\eta/s=0.2$ and $\chi=3.0$.}
\label{fig:FOSurf1}
\end{figure}

A typical freezeout surface for $\chi=3$ at an impact parameter b=6.5
is shown in fig.~\ref{fig:FOSurf1}.  Color gradients show the
temperature profile on the freezeout surface and as noted before the
surface is not necessarily an isotherm.

\begin{figure}[hbtp]
  \vspace{9pt}
  \centerline{\hbox{ \hspace{0.0in} 
\includegraphics[scale=0.725]{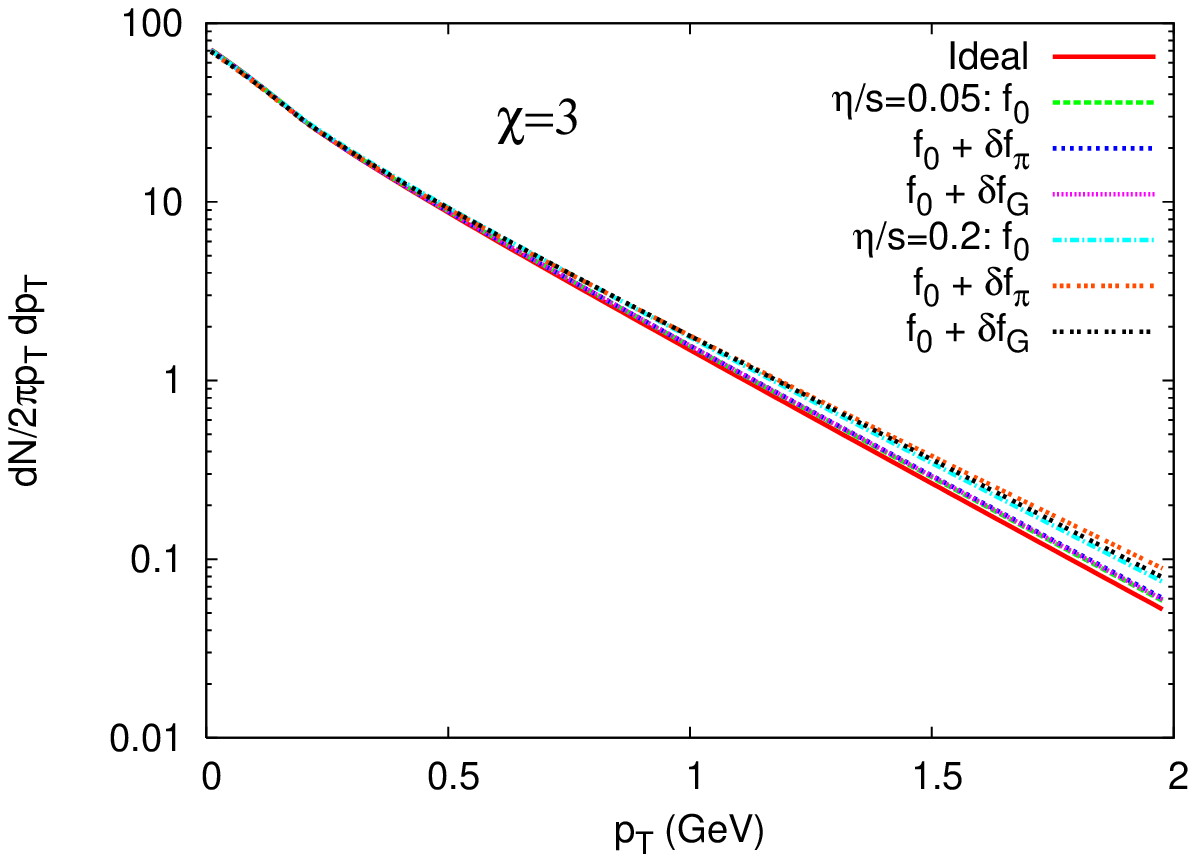}
    \hspace{0.0in}
\includegraphics[scale=0.725]{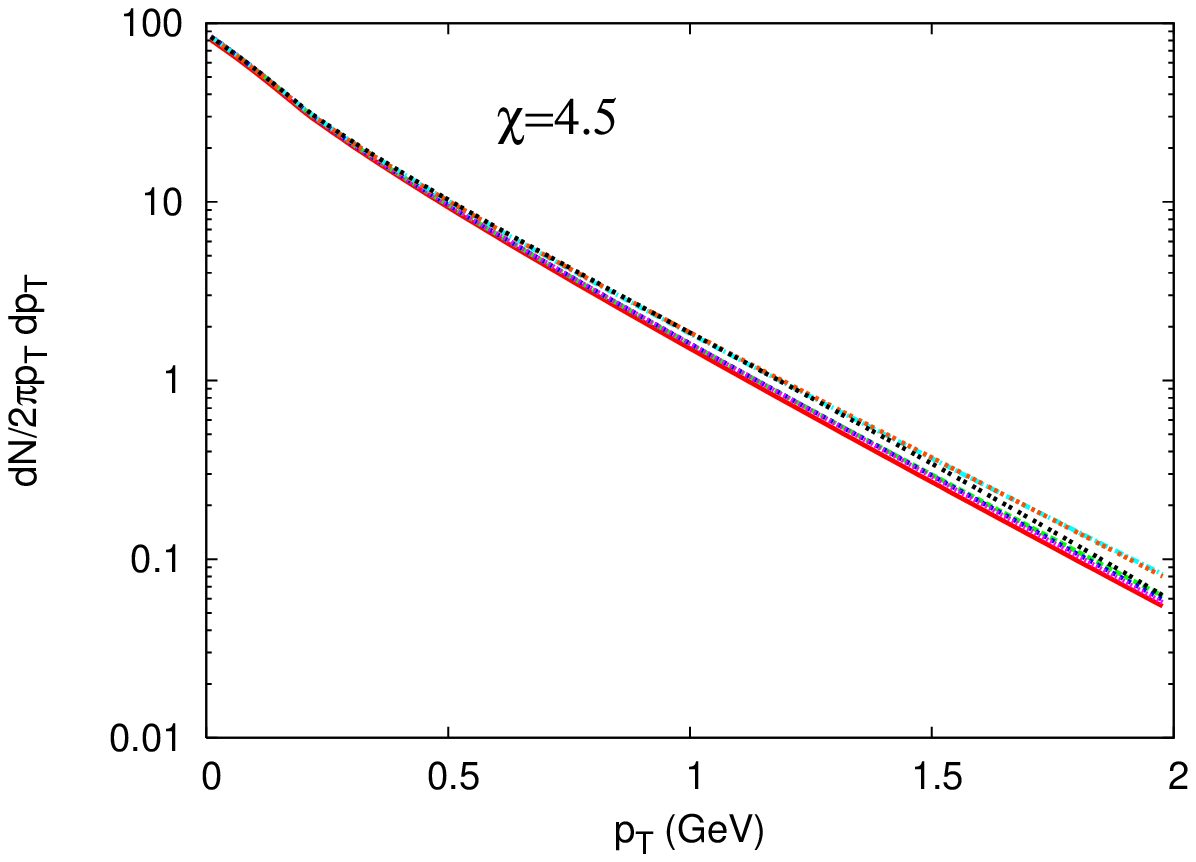}
    }
  }
\caption{(Color online) Differential transverse momentum spectra for Au-Au collisions at b=6.5 fm.  The left plot is for freeze-out parameter $\chi=3$ and the right for $\chi=4.5$.  In both plots the ideal case is shown by the solid red curve.  Then the viscous case is shown without including the viscous corrections to the distribution function and is denoted by $f_o$.  The addition of the viscous correction to the distribution function is generated in two different ways.  $\delta f_\pi$ is calculated using the auxiliary tensor $c^{\mu\nu}$ while $\delta f_G$ is calculated using the physical gradients {\em i.e., $\pi^{\mu\nu}=-\eta\langle \partial^\mu \partial^\nu\rangle$}.    }
\label{fig:ptsp}
\end{figure}

Differential $p_T$ spectra for massless particles are shown in
fig.~\ref{fig:ptsp} for two different freezeout surfaces: $\chi=3.0$
(left) and $\chi=4.5$ (right).  In both plots the ideal case is shown
by the solid red line.  First we discuss changes to the spectra
brought about by modifications to the equations of motion by looking at
the spectra generated with the ideal particle distribution ($f_o$
only).  For both values of viscosity and both freezeout choices a
hardening of the spectra is observed.  This is expected since viscosity
tends to increase the transverse velocity. 

The effect from the viscous corrections to the distribution function
are more subtle.  At earlier times the transverse flow has not fully
developed and the longitudinal pressure is reduced while the 
transverse pressure is increased \cite{Teaney:2003kp}. This 
is a consequence of the fact that the shear tensor is traceless.
The increase in transverse pressure 
leads to a hardening of the spectrum after integration
over the space-time freezeout surface. 
This is the case for $\chi=3$
even though the corrections are small.  At later times the larger 
transverse
flow alleviates some of the longitudinal shear. When the hydro is
finally in a full 3D expansion, the viscous correction tends to 
reduce the transverse pressure. This 
changes the sign of the viscous correction term.
This is seen for $\chi=4.5$ where the viscous corrections soften the
spectrum slightly. 

As discussed above, any observable created by using the auxiliary
variable $c^{\mu\nu}$ should agree with the results using the physical
velocity fields.  Therefore we  also show the viscous corrections
calculated using the physical gradients (denoted by $\delta f_G$), {\em i.e.,}
in the local rest frame the $\pi^{ij}$ is approximated by
\st
    \pi^{ij} = -\eta(\partial^i u^j + \partial^j u^i - \frac{2}{3}\delta^{ij} \partial_l u^{l})\, , 
\stp 
when computing $\delta f$.

Overall, the corrections to the spectra are small so it is hard to see
any differences between the two calculations.  This will not be the
case for the differential elliptic flow where this comparison will be
more important.

\begin{figure}[hbtp]
  \vspace{9pt}
  \centerline{\hbox{ \hspace{0.0in} 
\includegraphics[scale=0.725]{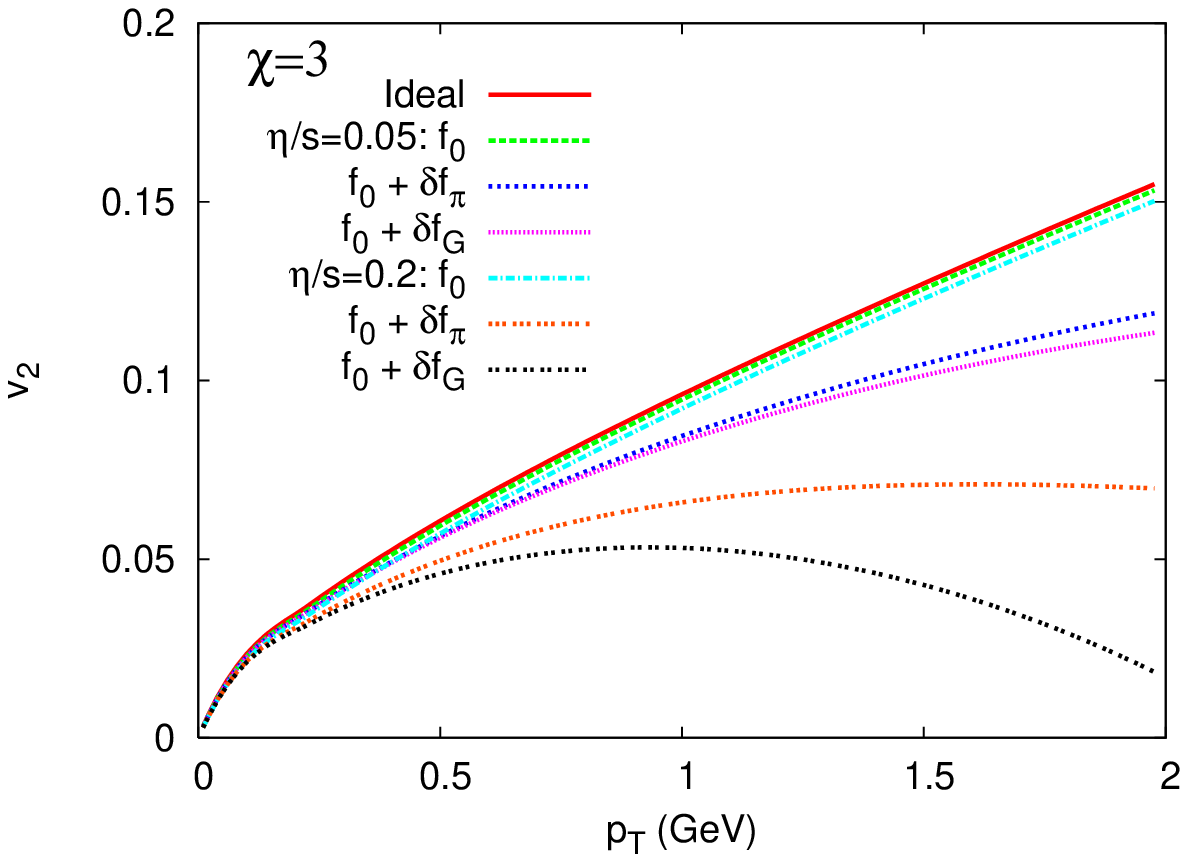}
    \hspace{0.0in}
\includegraphics[scale=0.725]{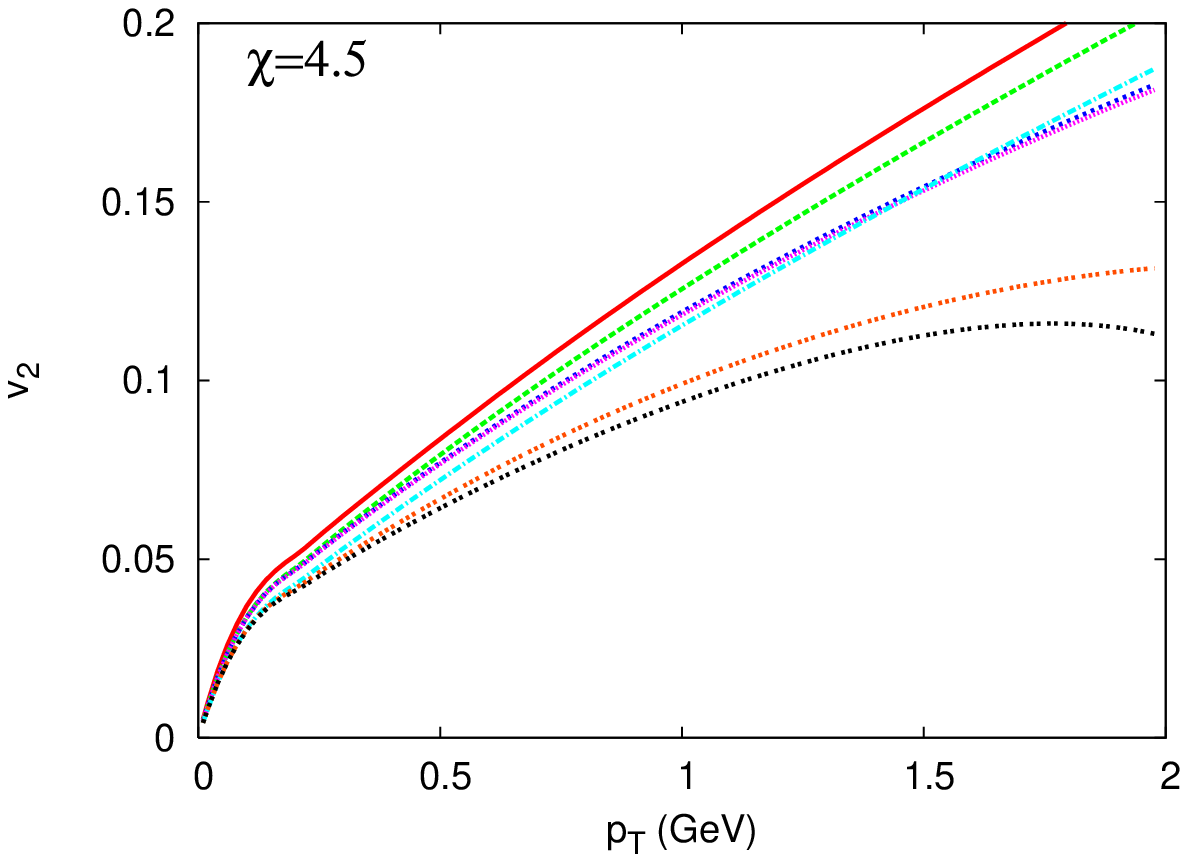}
    }
  }
\caption{(Color online) Differential $v_2$ spectra for Au-Au collisions at b=6.5 fm.  The resulting curves are generated in the same way as described for the $p_T$ spectra in fig.~\ref{fig:ptsp}}
\label{fig:v2sp}
\end{figure}

Figure~\ref{fig:v2sp} shows the differential elliptic flow using the
same parameter set from the $p_T$ spectrum.  The solid red curves
shows the ideal spectrum and, as expected, a larger elliptic flow is
generated for $\chi=4.5$ compared to $\chi=3$ since a larger fraction
of the space-time volume is described by hydrodynamics.

The viscous correction to the equations of motion causes only a small
change in the elliptic flow as seen by comparing the results at finite
viscosity using $f_o$ only with the ideal case.  For $\chi=3$ the
change is almost negligible.  For $\chi=4.5$ deviations are on the
order of 2\% at $p_T=2$ GeV.   

Including the viscous corrections to the distribution function can
bring about large changes in the elliptic flow.  We show the
corrections due to the auxiliary variable by $\delta f_\pi$ and those
from the gradients by $\delta f_G$ and we expect the two results to
agree.  When the two results start to diverge the gradient expansion is
no longer valid and a kinetic description is really required.

Based on our discussion in section \ref{sec:fo} the viscosity is what
sets the freezeout surface.  For $\eta/s=0.2$ we find that $\chi=3$
(figure on left).  In this case the viscous corrections are large but
can only be trusted up to $p_T \approx 1$ GeV.  We also show for
comparison the spectra for $\eta/s=0.05$ which can be trusted past 2
GeV.  For $\eta/s=0.05$ we take $\chi=4.5$ for reasons discussed in
section \ref{sec:fo}.  Again, the viscous correction decreases the
elliptic flow as a function of $p_T$.  Also shown are the spectra for
$\eta/s=0.2$ and the corrections are larger. In both cases the spectra
can be trusted past $p_T$ = 2 GeV.   

\section{Discussion and comparison with other works}

\subsection{Discussion}

In summary we now make several conclusions regarding the effects of
shear viscosity on heavy ion collisions.  

We first recall the setup. The paper is restricted to an ideal gas
equation of state $p=\frac{1}{3}\epsilon$  and sets the initial
non-equilibrium fields to the value expected from the navier stokes
equations $\pi^{ij} = -\eta \llangle \partial^i u^j\rrangle$. The
initial distribution of entropy density follows the distribution of
participants. (This could be changed to a Color Glass Condensate model
initial conditions \cite{Hirano:2005xf}.) The paper simulates a fluid model based on \cite{Ottinger}
which is similar but differs from that of Israel and Stewart.  However
all models should ultimately agree on the magnitude of viscous
corrections provided the viscosity is sufficiently small.  

Several technical notes warrant discussion here. An algorithm for
a reliable solution of the viscous model was developed by Pareschi \cite{Pareschi} and is presented in
appendix~\ref{sec:Alg} which achieves uniform numerical accuracy
across a wide range of relaxation times.  For small enough relaxation
times the auxiliary fields $\pi^{ij}$ should relax to the form
expected from the Navier-Stokes equation $\pi^{ij}\simeq-\eta \llangle
\partial^i u^j \rrangle$. To see this
good/reasonable convergence for small/modest viscosities see
appendix~\ref{sec:grad}. Generically, relaxation models for viscosity
have long time parameters (the shear viscosity $\eta$ in this case)
and short time parameters. In the model considered here,
$\alpha$ (see appendix~\ref{sec:a}) is the 
short time parameter while  in the 
Israel-Stewart theory this short time parameter is $\eta/[(\epsilon +
p) \tau_{\pi}]$. These short time parameters can be constrained by the
$f$-sum rule \cite{KadanoffMartin, Teaney_corr, PTcorr} and is discussed further in appendix~{\ref{sec:a}}. In
general the results should not depend on these short time parameters.

We now summarize our physical results.  The integrated viscous
corrections to the flow are small.  This was seen in both the
hydrodynamic fields and also in the differential and integrated
elliptic flow when the thermal distribution function was restricted to
the ideal form.  (The remainder of this paragraph discusses only 
results with this restriction.) For the integrated $v_2$ this is seen in
the left plot of fig.~\ref{fig:A2} where $A_2$ is shown for ideal runs and
viscous runs at $\eta/s=0.05$ and $0.2$. Corrections due to the
modified flow pattern are also small in the  differential $v_2$
spectrum as seen in fig.~\ref{fig:v2sp} by comparing the ideal and
viscous runs (again with $f_o$ only.) Although there is the possibility
for the elliptic flow to be modified from variations in the freezeout
surface across different runs this was minimized by freezing out on
contours of constant $\chi$.  One can see from fig.~\ref{fig:fosurf}
that the space-time freezeout contours are about the same at zero and
finite viscosity.  The fact that only small changes in the fields are
seen when including viscosity is not surprising.  The time scale of any
heavy ion collision is much shorter then the time needed for
dissipative effects to integrate and become large. 

\begin{figure}[hbtp]
  \vspace{9pt}
  \centerline{\hbox{ \hspace{0.0in}
\includegraphics[scale=0.75]{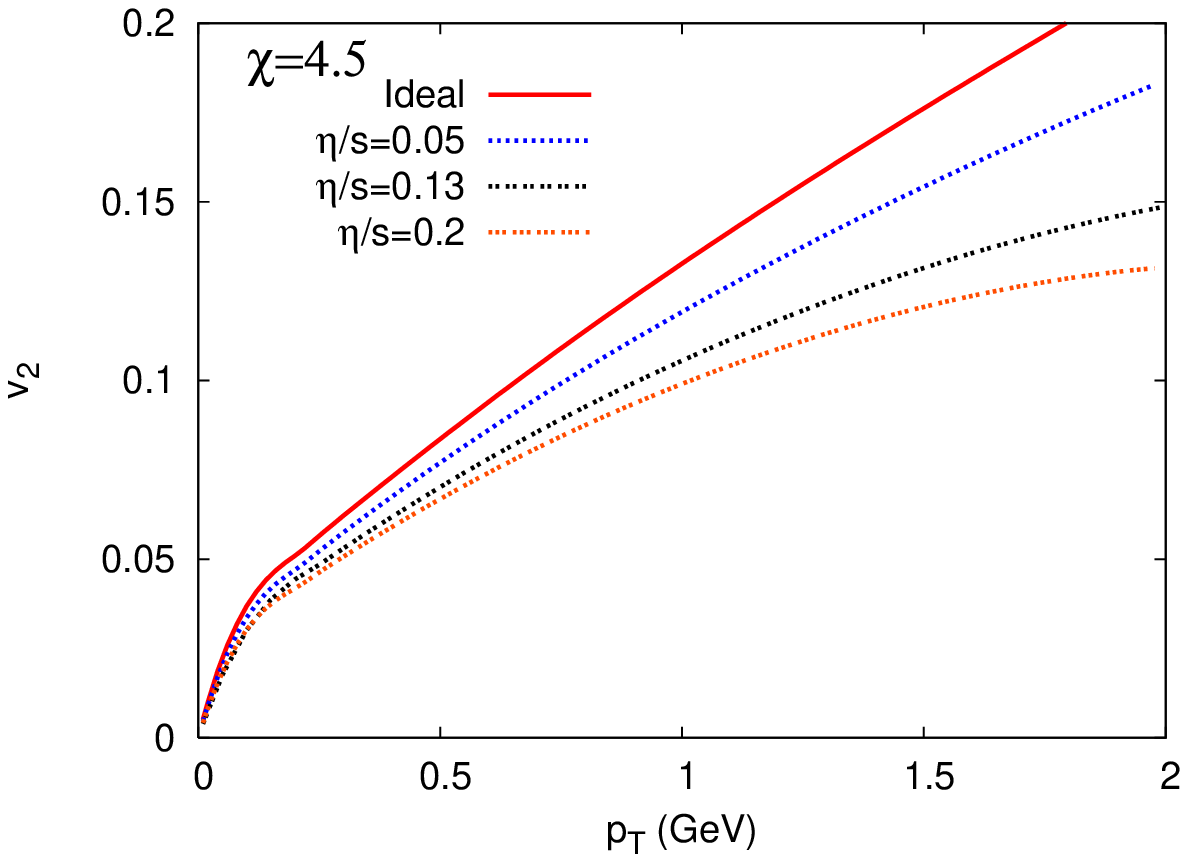}
    \hspace{0.0in}
\includegraphics[scale=0.75]{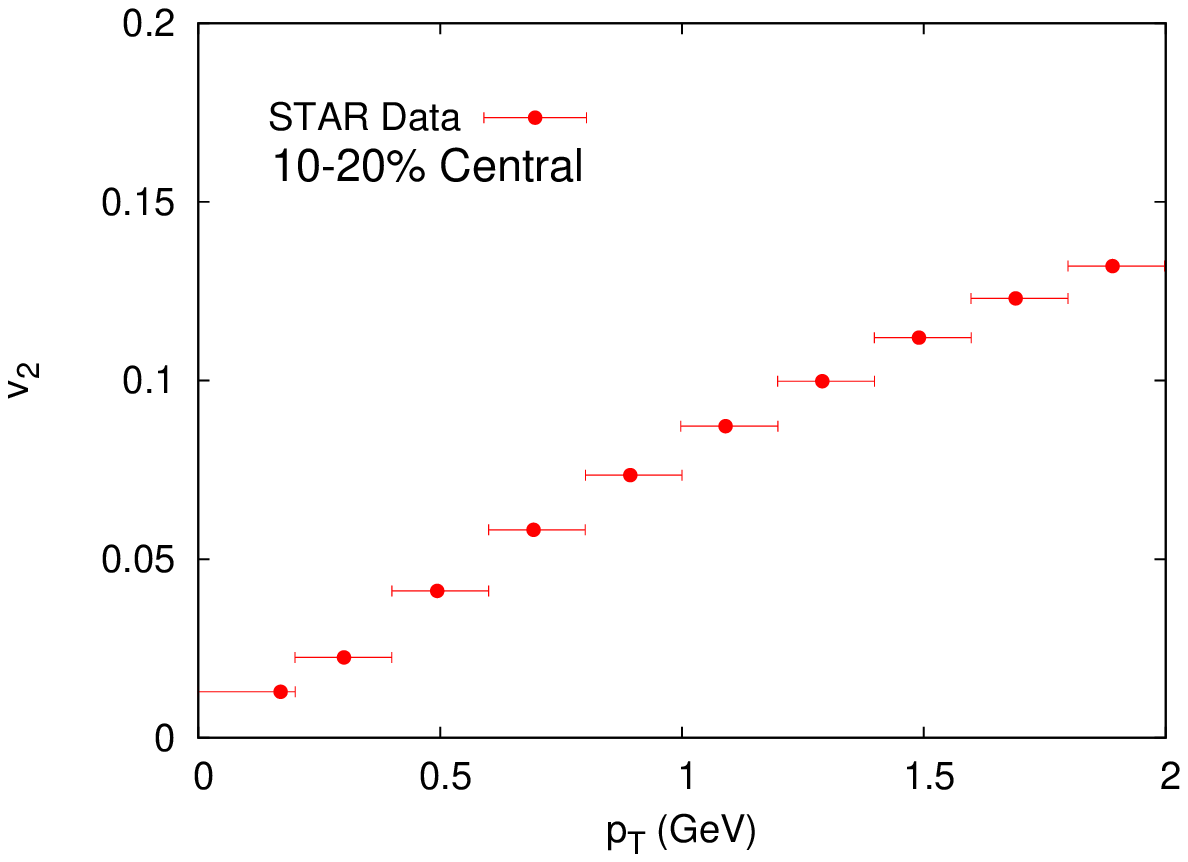}
    }
  }
\caption{Left: Summary plot showing $v_2$ for massless particles for simulations using ideal hydro and $\eta/s=0.05, 0.2$.  Right: Charged hadron $v_2$ data using the standard reaction plane method as measured in Au-Au collisions at $\sqrt{s}=200$ GeV for a centrality selection of 10\% to 20\% \cite{Adams:2004bi}.}
\label{fig:v2sum}
\end{figure}

Even though viscosity does not modify the flow strongly we have shown
that there are still large corrections to the particle spectra due to
off-equilibrium corrections to the ideal particle distribution
function.  Any bounds for the viscosity (at least from this paper)
would have to come from the $v_2$ spectra.  As Lindblom \cite{Lindblom}
and earlier work by others \cite{KadanoffMartin}  has clarified, any
observable computed from the auxiliary fields $\pi^{ij}$ must agree
with the same observable generated by the physical gradients
$-\eta\llangle \partial^i u^j\rrangle$. When
deviations are seen  the viscous corrections can no longer be  trusted.
For a freezeout surface set by $\chi=4.5$ the viscous corrections agree
with gradients up to 2 GeV for viscosities as large as $\eta/s=0.2$ as
seen in figure \ref{fig:v2sp}.  It is therefore safe to use only the
auxiliary variable when generating spectra for this particular
parameter set.  In figure~\ref{fig:v2sum} we show a summary plot of the
differential elliptic flow.  We now show one additional curve for
$\eta/s=0.133$ yielding $(\eta/p)\partial_\mu u^\mu=0.6$ for this
particular choice of  freezeout surface.  We believe that this choice
of parameters is the closest physical scenario.  The right plot of
figure \ref{fig:v2sum} shows the measured elliptic flow 
as measured by the STAR collaboration \cite{Adams:2004bi}.  We do not intend to
make a comparison, but simply would like to keep the data in mind.
Nevertheless since this simulation was performed with a massless gas
which has the largest elliptic flow, it seems difficult to imagine that
the $\eta/s \gsim 0.35$ will ever fit the data even if the initial 
conditions are modified along the lines of Ref.~\cite{Hirano:2005xf}.

Before a realistic comparison with data can be made the QGP/hadronic
phase transition must be taken into account.  In the vicinity of the
phase transition it is possible that the shear viscosity may become
very large.  Also, a more realistic model for the hadronic gas would be
the hard sphere model where $\eta\sim\frac{T}{\sigma_0}$.  This would
adjust at what point the simulation freezes out and would therefore
effect spectrum.  There is most likely a finite bulk viscosity due to
the fluctuations of the QGP and hadron concentrations in the mixed
phase or from chemical off-equilibrium in the hadronic phase
\cite{Kharzeev:2007wb}.  A final issue that should be taken into consideration
is that particles of different mass could possibly freezeout 
on different surfaces.  These issues will be addressed in a
future work.

\subsection{Comparison}
\label{sec:compare}

We now compare our results to some other groups, first with the recent
results of Song and Heinz \cite{Heinzv2} where they computed
differential $v_2$ spectrum in Cu-Cu collisions.  

One conclusion they found is that varying the initial conditions do not
change the end result, even in the extreme condition when the
equilibrium stress tensor is set to zero $\pi^{\mu\nu}(\tau_0)=0$.  
This insensitivity is similar to the insensitivity to the short 
time parameter $\alpha$ indicated in appendix~\ref{smalltime}.

Song and Heinz also found that the viscosity substantially changes the
flow.  Their differential $v_2$ spectra changes
dramatically when viscosity is included, even if the particles
freezeout using an ideal distribution.  In our case there is almost no
change in $v_2$ when viscosity is present when freezing out with $f_o$
only, while in their case they see $v_2$ decrease by a factor of two at
3 GeV due to changes in the flow alone (see their fig. 4).  It is possible that this
difference is due to their inclusion of a phase transition in their
equation of state.  Once the plasma phase reaches the phase transition
the momentum anisotropy ($\epsilon_p$) stalls in their model staying
constant for the entire mixed phase (and only growing slightly in the
hadronic phase).  Therefore after freezing out the elliptic flow is in
some sense probing earlier times where viscous corrections to the flow
may be larger.  Actually, if one looks at our $\epsilon_p$ in figure
\ref{fig:anis} the largest differences between the ideal and viscous
cases is for earlier times. This explanation should be verified. 

When the viscous corrections to the distribution function are added we both see qualitatively the same behavior.  The viscous correction gets larger with transverse momentum eventually driving the elliptic flow below zero.  It is impossible to make quantitative comparisons until we include a phase transition and run simulations with smaller systems sizes, which we plan on doing in a future work.   




We now compare our work with that of Baier and Romatschke
\cite{Baier:2006gy, Romatschke:2007jx} for the flow and
$p_T$ spectrum in central collisions.  They also find that the
viscosity does not integrate to significantly modify the ideal flow.
We both find qualitatively the same behavior confirming even earlier works \cite{Muronga,DT1}.  Finite
viscosity causes the temperature to drop slowly at earlier times
and more quickly at later times compared to the ideal case.
This effect was already discussed in section~\ref{sec:HydroResults}  

Although  the  freeze-out conditions in this work differ from those 
of Baier and Romatschke,  we find qualitatively the same behavior when comparing $p_T$ spectra, {\em i.e.} a hardening of spectra at large $p_T$. 

In comparison with the differential $v_2$ results by Romatschke and
Romatschke \cite{RRv2} we see qualitatively the same behavior.  In this
case the comparison is more direct  since both simulations were
performed using Au-Au collisions at the same $\sqrt{s}$.   Comparing
our results for $\chi=4.5$,  we see that our $v_2$ drops by
$\approx$50\% at $p_T=2$ GeV.  This is on the same order as
seen in fig.~3 of Ref.~\cite{RRv2}. However they do not show
contributions from flow and the distribution function separately so it
is hard to make any definitive comparisons.

\section{Conclusions}

In this work we have outlined the equations of motion necessary for a casual description of viscous relativistic hydrodynamics and have shown results using initial conditions tuned to Au-Au collisions at $\sqrt{s}\approx 200$ GeV.  The results indicate that the viscous correction to the ideal equations of motion are small.  The goal of this work was to calculate the viscous correction to differential $v_2$ spectra.  Even though modifications to $v_2$ from the flow are small the effect of the off-equilibrium distribution function can bring about large changes in $v_2(p_T)$.  By requiring observables calculated with the auxiliary fields to agree with those calculated with the physical gradients one can identify where a hydrodynamic description is reliable.

\appendix

\section{Sphericity}
\label{sec:sph}
In this appendix we show (following \cite{OllitraultSph}) how the momentum anisotropy, expressed through $A_2$, can be related to hydrodynamic quantities.  As discussed in the text $A_2$ is defined as
\bg
A_2=\frac{S_{11}-S_{22}}{S_{11}+S_{22}}=\frac{\langle p_x^2\rangle-\langle p_y^2\rangle}{\langle p_x^2\rangle+\langle p_y^2\rangle}.
\nd

The sphericity tensor $S_{ij}$ is calculated from the third moment of the momentum distribution function
\bg
S_{ij}=\int_\sigma d\sigma_\mu S^{\mu\nu\rho},
\nd
where $d\sigma_\mu$ is a differential element of the freezeout surface and the third-rank tensor, $S^{\mu\nu\rho}$ is defined as
\bg
S^{\mu\nu\rho}=\int p^\mu p^\nu p^\rho f(p)\frac{d^3p}{(2\pi)^3E_\p}.
\nd

In order to relate the sphericity tensor to hydrodynamic quantities we follow the same steps as was done in \cite{OllitraultSph} but also include the additional terms coming from viscous corrections.  First one substitutes the expression for the momentum distribution function with the appropriate viscous correction term into the above equation for the third-rank sphericity tensor
\bg
S^{\mu\nu\rho}=S_I^{\mu\nu\rho}+S_V^{\mu\nu\rho}.
\nd
The subscripts $I$ and $V$ correspond to the ideal and viscous contributions respectively and are defined as
\bg
S_I^{\mu\nu\rho}=\int p^\mu p^\nu p^\rho f_o(p)\frac{d^3p}{(2\pi)^3E_\p},\\
S_V^{\mu\nu\rho}=\frac{1}{2sT^3}\left[\pi_{\alpha\beta} + \frac{2}{5}\Pi\Delta_{\alpha\beta}\right] S_5^{\mu\nu\rho\alpha\beta},
\nd
where we have defined the fifth-rank tensor
\bg
S_5^{\mu\nu\rho\alpha\beta}=\int p^\mu p^\nu p^\rho p^\alpha p^\beta f_o(p)\frac{d^3p}{(2\pi)^3E_\p}.
\nd
Lorentz invariance sets the form of both $S_I$ and $S_5$ as follows
\bg
S_I^{\mu\nu\rho}=Au^\mu u^\nu u^\rho+B(g^{\mu\nu} u^\rho + \text{permutations})\\,
S_5^{\mu\nu\rho\alpha\beta}=Cu^\mu u^\nu u^\rho u^\alpha u^\beta +D(u^\mu u^\nu u^\rho g^{\alpha\beta} + \text{permutations})+\nonumber\\E(u^\mu g^{\nu\rho}g^{\alpha\beta} + \text{permutations}).
\nd

The coefficients A..E can be found in the same manner as was done previously in \cite{OllitraultSph}.  Quoting the result:
\bg
A=n(2\dlangle E^2\drangle-m^2) \\
B=n\frac{\dlangle E^2\drangle-m^2}{3} \\
C=n(16\dlangle E^4\drangle-16m^2\dlangle E^2\drangle+3m^4)/3\\
D=-n(8\dlangle E^4\drangle-11m^2\dlangle E^2\drangle+3m^4)/15\\
E=n(\dlangle E^4\drangle-2m^2\dlangle E^2\drangle+m^4)/15
\nd

The results for $S^{11}_I$ and $S^{22}_I$ can be expressed in terms of hydrodynamic quantities:
\bg
S^{11}_I=\int_\sigma (A u_x^2+B) u^\mu d\sigma_\mu+2B \int_\sigma u^x d\sigma_x \nonumber\\
S^{22}_I=\int_\sigma (A u_y^2+B) u^\mu d\sigma_\mu+2B \int_\sigma u^y d\sigma_y
\nd

The result for $S^{11}_V$ and $S^{22}_V$ can be found making use of the following three identities:
\bg
u^\alpha\pi_{\alpha\beta}=0 \nonumber\\
g^{\alpha\beta}\pi_{\alpha\beta}=0 \nonumber\\
u^\alpha\Delta_{\alpha\beta}=0
\nd
Therefore, the viscous correction to the sphericity tensor is given as
\bg
S^{11}_V=\frac{1}{2sT^3}\int_\sigma 2E\left[ \pi^{xx} u^\mu d\sigma_\mu+2u^x\pi^{x\mu} d\sigma_\mu \right]+\nonumber\\
\frac{\Pi}{5sT^3}\int_\sigma\left[ 3D u_x^2 u^\mu d\sigma_\mu+E\left( 5u^\mu d\sigma_\mu+10u^x d\sigma_x-6 u_x^2 u^\mu d\sigma_\mu\right)\right] ,
\nd
and
\bg
S^{22}_V=\frac{1}{2sT^3}\int_\sigma 2E\left[ \pi^{yy} u^\mu d\sigma_\mu+2u^y\pi^{y\mu} d\sigma_\mu \right]+\nonumber\\
\frac{\Pi}{5sT^3}\int_\sigma\left[ 3D u_y^2 u^\mu d\sigma_\mu+E\left( 5u^\mu d\sigma_\mu+10u^y d\sigma_y-6 u_y^2 u^\mu d\sigma_\mu\right)\right].
\nd

\section{Viscous correction to the distribution function}
\label{sec:vc}

The thermal $p_T$ and differential $v_2$ spectra of particles are
generated using the Cooper-Frye formula \cite{CF}
\st
E\frac{d^3N}{d^3p}=\frac{g}{2\pi^3}\int_\sigma f(p_\mu u^\mu,T) p^\mu d\sigma_\mu,
\stp
where $d\sigma_\mu$ is the normal vector to the freezeout surface set
by the condition of constant $\chi$.  For the geometry we are considering here we have
\bg
p_\mu u^\mu=m_T u^0 \cosh(\eta_s)-p_1 u^1-p_2 u^2, \\
p^\mu d\sigma_\mu = \tau( m_T \cosh(\eta_s) d\sigma_0 + p^1 d\sigma_1 + p^2 d\sigma_2 ).
\nd

Following \cite{Teaney:2003kp,AMY} we will make a second moment 
ansatz for the thermal distribution function. We first write the 
stress tensor as 
\st
   T^{\mu\nu} = \epsilon u^{\mu} u^{\nu} + (p + \Pi) \Delta^{\mu\nu} + \pi^{\mu\nu},
\stp
where  $\pi^{\mu\nu}$ is symmetric traceless and satisfies 
$\pi^{\mu\nu}u_{\nu}=0$. 
Then we subsequently make an ansatz for the thermal
distribution  $f \rightarrow f_o + \delta f$
\st
   \delta f = \frac{1}{(e + p) T^2}\, f_o(1 + f_o) \, 
p^{\mu} p^{\nu} \left[
\frac{C_1}{2} \pi_{\mu\nu} + \frac{C_2}{5} \Pi \Delta_{\mu\nu}  \right]
\stp
where $C$ is a constant and 
the factor $1/\left[(e + p)  T^2\right]$ has been 
inserted for later convenience. 
To determine the constant $C$ we demand that
\st
   T^{\mu\nu}= \int \frac{d^3p}{(2\pi)^3} \, \frac{p^{\mu}p^{\nu}}{E_{\p} } \, f.
\stp
Working in the local rest frame this becomes a condition that 
\st
 \Pi \delta^{ij} + \pi^{ij} = \left\{\frac{1}{(e+p) T^2}\, \int \frac{d^3p}{(2\pi)^3} \,\frac{p^i p^jp^l p^m}{E_\p} \,  f_o(1 + f_o)\,\right\} \left[\frac{C_1}{2} \pi_{lm} + \frac{C_2}{5} \Pi \delta_{lm} \right].
\label{eq:c1c2}
\stp
The integral over the three momentum in curly braces can be expressed as
\st
 I\left(\delta^{ij}\delta^{lm} + \delta^{il}\delta^{jm} + \delta^{im} \delta^{jl} \right),
\label{eq:6delta}
\stp
where
\st
 I = \frac{1}{15 (e + p) T^2 } \int \frac{d^3p}{(2\pi)^3} \, \frac{\left|\p\right|^4 }{E_\p} \, f_o ( 1 + f_o).
\stp
Inserting \Eq{eq:6delta} into \Eq{eq:c1c2} we see that  
\st
    C_1 = C_2 = \frac{1}{I}.
\stp 

We record two limiting cases of this integral.  In the massless limit with zero chemical potential the
integral is easily performed and yields
\bg
  I &=&  \frac{90 \zeta(5)}{\pi^4}\approx 0.958.
\nd
In the classical limit the factor $f_o(1+f_o)$  is 
replaced by
\st
 f_o(1 + f_o) \rightarrow f_o = e^{{-(E_\p-\mu)}/T},
\stp
and the integral is easily performed using the integral 
representation of the modified Bessel functions
\bg
 I &=& \frac{1}{(e + p) T^2}\, \left[\frac{m^3 T^3}{2\pi^2}\, e^{\mu/T} K_{3} (m/T)\right] = 1.
\nd

\section{Relaxation Time}
\label{sec:a}

It was shown in \cite{Teaney_corr, PTcorr} that for a weakly interacting theory the transport time scale $\tau_R$ is much longer than the inverse temperature.  This separation of time scales is seen in the spectral density which will have a sharp peak at small frequencies $\omega\sim1/\tau_R\ll T$.  In \cite{Teaney_corr, PTcorr} a sum rule for this peak was derived relating the small time ($t\ll \tau_R$) behavior of hydrodynamic correlators to the microscopic time scale.  The statement of this sum rule can be written as:
\bg
\frac{T}{k^2}\partial_t\chi_{gg}^L({\bf k},t)\vert_{t\sim1/\Lambda}\approx T(\epsilon+p)\langle\frac{3}{5}v_{{\bf p}}^2\rangle \nonumber\\
\frac{T}{k^2}\partial_t\chi_{gg}^T({\bf k},t)\vert_{t\sim1/\Lambda}\approx T(\epsilon+p)\langle\frac{v_{{\bf p}}^2}{5}\rangle \nonumber\\
\label{eq:sumrule}
\nd
where $v_{{\bf p}}$ is ${\bf p}/E$ and $\Lambda$ is a cut-off such that $1/\tau_R\ll\Lambda\ll T$. $\chi_{gg}$ is the retarded correlator of $T^{0i}$ and can be found in the framework of linear response theory.  A small velocity field is turned on with a perturbing Hamiltonian
\bg
H=H_0-\int d^3{\bf x} v^i({\bf x},t)T^{0i}({\bf x},t),
\nd
and suddenly switched off at t=0: $v^i({\bf x},t)=e^{\epsilon t}\theta(-t)v_0^i({\bf x})$.  In the framework of linear response this yields
\bg
\partial_t\langle T^{0i}({\bf k},t)\rangle=-\chi_{gg}^{ij}({\bf k},t)v_0^i({\bf k}).
\nd 

The stress tensor can be expressed as the equilibrium stress tensor plus small corrections:
\bg
\langle T^{00} \rangle = e+\epsilon({\bf x},t) \\
\langle T^{0i} \rangle =0+g^i({\bf x},t), \\
\nd
where ${\bf g}\equiv{\bf v}(e+p)$.  The linearized hydrodynamic equations are:
\bg
\partial_t \epsilon+\partial_ig^i=0\\
\partial_t g^j+\partial_i \tau^{ij}=0,
\label{eq:linhyd}
\nd
where
\bg
\tau^{ij}=\delta^{ij}p-\eta(\partial^i u^j+\partial^j u^i-\frac{2}{3}\delta^{ij}\partial_l u^l)-\delta^{ij}\zeta\partial_l u^l,
\nd
in the Navier-Stokes limit.  However, since we are interested in relating the short time parameters of the theory used in this work to microscopic quantities we take $\tau^{ij}$ from the second-order equations
\bg
\tau^{ij}=p(\delta^{ij}-\alpha c^{ij}).
\nd

We now have all the pieces needed in order to evaluate the left hand side of eq.~\ref{eq:sumrule}.  First, for small $c$ the evolution equation simplifies to
\bg
\partial_t c^{ij}-(\partial^i u^j+\partial^j u^i)=\frac{1}{\tau_0}c^l_l\delta^{ij}+\frac{1}{\tau_2}(c^{ij}-\frac{1}{3}c^l_l\delta^{ij}).
\nd

We can now differentiate  eqn.~\ref{eq:linhyd} with respect to time, substitute in the evolution equation and immediately take the $t\to0$ limit in order to obtain
\bg
\partial_t^2 g^j\vert_{t=0}=-c_s^2\partial_j\partial_t\epsilon-\alpha p \partial_i[\partial^i u^j+\partial^j u^i].
\label{eq:relax1}
\nd

We now make use of the first linearized hydrodynamic equation, $\partial_t\epsilon=-(\epsilon+p)\partial_i v^i$, which can be substituted into eq.~\ref{eq:relax1}.  After taking a spatial Fourier transform we get:
\bg
\partial_t^2 g^j\vert_{t=0}=-c_s^2(\epsilon+p)k_j k_i v^i-p\alpha[k^2v^j+k^i k^j v^i].
\nd

This equation can be decomposed into its transverse and longitudinal pieces by defining $g^j=g^{jT}+\frac{k^j}{k}g^L$ where $k^jg^{jT}=0$.  Since we are only interested in the shear viscosity we can simply look at the transverse component
\bg
\partial_t^2g^{jT}=-p\alpha k^2v^j,
\nd

and when substituted into the sum rule, eqn.~\ref{eq:sumrule}, we obtain the result
\bg
\alpha=\frac{4}{5}.
\nd

A similar analysis can be done for the Israel-Stewart equations \cite{IS} as well:
\bg
\tau^{ij}=p\delta^{ij}+\pi^{ij}\\
\partial_t\pi^{ij}=\frac{1}{\tau_\pi}[-\eta\langle\partial^i v^j\rangle-\pi^{ij}]
\nd

with the result
\bg
\tau_\pi=\frac{5\eta}{4p}.
\nd

\subsection{Dependence on small time parameter}
\label{smalltime}

\begin{figure}[hbtp]
  \vspace{9pt}
  \centerline{\hbox{ \hspace{0.0in} 
\includegraphics[scale=0.725]{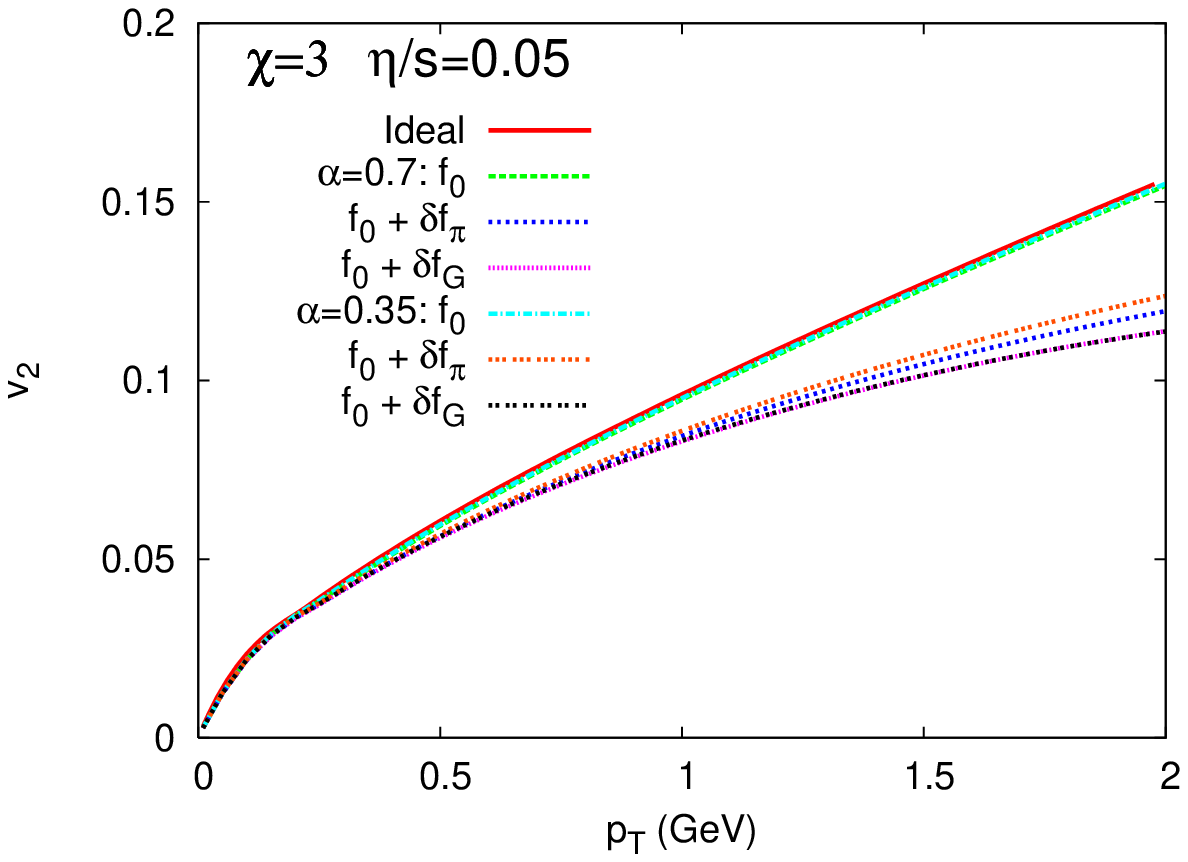}
    \hspace{0.0in}
\includegraphics[scale=0.725]{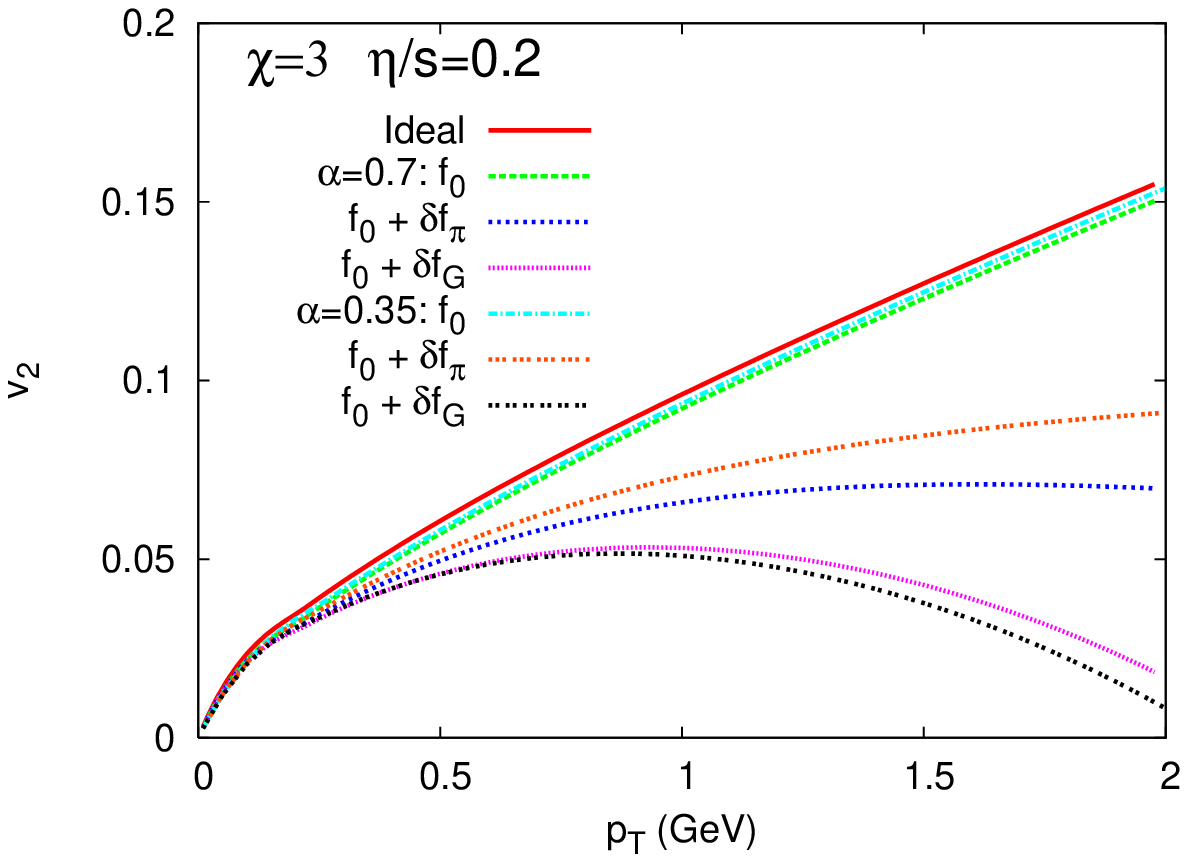}
    }
  }
\caption{Differential $v_2$ spectrum for Au-Au collisions at b=6.5 fm.  Left: $\eta/s=0.05$ Right: $\eta/s=0.2$ Each figure shows spectrum calculated with the default small time parameter $\alpha=0.7$ and half of this value, $\alpha=0.35$. }
\label{fig:a1dep}
\end{figure}

It was discussed throughout the paper that the results should not depend on the small time parameter.  In order to test this we have generated $v_2$ spectrum with a value of $\alpha=0.35$ compared to the default value of $\alpha=0.7$ used throughout this work.  This is shown in Fig.~\ref{fig:a1dep} for $\eta/s=0.5$ (left) and $\eta/s=0.2$ (right) for a fixed freeze-out surface parameter $\chi=3$.  The ideal curves are also shown for reference.

The deviations between the results using two different values of $\alpha$ are small.  The flow is hardly changed as seen by comparing the spectrum generated using only the ideal distribution function, $f_o$.  In this case, for both values of $\eta/s$, the results differ by less than 1\%.  When including the viscous correction to the spectrum the results still agree reasonably well.

\section{Algorithm}
\label{sec:Alg}

In this appendix the algorithm used in order to solve Eqn's~\ref{eq:2dstress}-\ref{eq:2dcs} is outlined.  The numerical evaluation of the above system of hyperbolic equations is difficult because one would like would like to achieve uniform numerical accuracy across a range of relaxation times.  In order to achieve this we use a discretization method first proposed by Pareschi \cite{Pareschi} which can numerically solve the above equations in both the stiff and unstiff regions.

We use notation such that the term $x^n_{i,j}$ refers to the value of x at discrete time $t^n$ and grid point $(x,y)=(x_i,y_j)$ with i and j always referring to the x and y grid coordinates.  Any variable absent of an index represents a continuous variable not yet specified at a given point.  At times we use a simplified notation for the discretized fields such that $u(x_i,y_j,t^n)=u^n_{i,j}$.

For completeness we outline the integration routine developed by \cite{Pareschi}.  Our goal is to solve equations of the form
\bg
u_t(x,y,t)+f_x(u)+h_y(u)=g(u),
\label{eq:Apde}
\nd
where u, f, h, and g are arbitrary functions.  We use the standard notation that for a finite volume element the value of a field at the point $(x_{i+\oh},y_{j+\oh})$ and at a time $t=t^{n}$ is given by:

\bg
u^{n}_{i+\oh,j+\oh}=\frac{1}{\Delta x\Delta y}\int^{x_{i+1}}_{x_i}\int^{y_{j+1}}_{y_j}u(x,y,t^{n})dx dy
\nd

We then integrate eq.~\ref{eq:Apde} over the region $[x_i,x_{i+1}]\times[y_j,y_{j+1}]\times[t^n,t^{n+1}]$ yielding:

\bg
u^{n+1}_{i+\oh,j+\oh}=\frac{1}{\Delta x\Delta y}\Bigglb[ \int^{x_{i+1}}_{x_i}\int^{y_{j+1}}_{y_j} u(x,y,t^n)dxdy+
\int^{y_{j+1}}_{y_j}\int^{t^{n+1}}_{t^n}\biglb( f(x_{i+1},y,t)-f(x_i,y,t)\bigrb)dydt+\nonumber\\
+\int^{x_{i+1}}_{x_i}\int^{t^{n+1}}_{t^n} \biglb(h(x,y_{j+1},t)-h(x,y_j,t)\bigrb) dxdt+
\int^{x_{i+1}}_{x_i}\int^{y_{j+1}}_{y_j}\int^{t^{n+1}}_{t^n} g(x,y,t)dxdydt\Biggrb]\nonumber\\
\label{eq:ap1}
\nd

The first integral over the field $u$ in the above equation (\ref{eq:ap1}) can be discretized by constructing a piecewise linear approximation of u(x,y,t) over the integration region:

\bg
\int^{y_{j+1}}_{y_j} u(x,y,t^n)dy = \int^{y_{j+\oh}}_{y_j} \left[ u(x,y_j,t^n)+(y-y_j)u_y(x,y_j,t^n)\right] dy+\nonumber\\
+\int^{y_{j+1}}_{y_{j+\oh}} \left[ u(x,y_{j+1},t^n)+(y-y_{j+1})u_y(x,y_{j+1},t^n)\right] dy
\nd

Using the corresponding linear approximation for the integration over $[x,x_{i+1}]$ and performing the elementary integration over $x$ and $y$ the following discretization is found:

\bg
\frac{1}{\Delta x\Delta y}\int^{x_{i+1}}_{x_i} \int^{y_{j+1}}_{y_j} u(x,y,t^n)dxdy = \frac{1}{4}\biglb( u^n_{i,j}+u^n_{i+1,j}+u^n_{i,j+1}+u^n_{i+1,j+1}\bigrb)+\nonumber\\
+\frac{\Delta x}{16}\partial_x\biglb( u^n_{i,j}-u^n_{i+1,j}+u^n_{i,j+1}-u^n_{i+1,j+1} \bigrb)+\frac{\Delta y}{16}\partial_y\biglb(u^n_{i,j}+u^n_{i+1,j}-u^n_{i,j+1}-u^n_{i+1,j+1} \bigrb)
\nd

For the time integrals over the fluxes (second and third term in eq.~\ref{eq:ap1}) a general trapezoidal rule is used:

\bg
\frac{1}{\Delta t}\int^{t^{n+1}}_{t^n} f(x,y,t)dt\approx \mu f(x,y,t^{n+\alpha})+ \nu f(x,y,t^n)
\label{eq:trapazoidgen}
\nd

where $f^{n+\alpha}$ will be given explicitly by a predictor step to be defined later.  The time integrals over the source term (last term in eq.~\ref{eq:ap1} will also be given by a general trapezoidal rule which will result in an implicit equation between the sources and charges.

\bg
\frac{1}{\Delta t}\int^{t^{n+1}}_{t^n} g(x,y,t)dt\approx \zeta g(x,y,t^{n+1})+ \eta g(x,y,t^n)
\label{eq:timesource}
\nd

In order to ensure second order accuracy in space the second and third integrals over the fluxes in eq.~\ref{eq:ap1} are evaluated after the time integrations using the standard midpoint rule and trapezoidal rule depending on the time:

\bg
\text{for }t=t^{n+1}:\text{     }\frac{1}{\Delta x}\int^{x_{i+1}}_{x_i} h(x,y,t)dx\approx h(x_{i+\oh},y,t)
\label{eq:midpoint}
\nd

\bg
\text{for }t\in [t^n,t^{n+1}]:\text{     }\frac{1}{\Delta x}\int^{x_{i+1}}_{x_i} h(x,y,t)dx\approx\frac{1}{2}\biglb[ h(x_{i+1},y,t)+ h(x_i,y,t)\bigrb]
\label{eq:trapezoid}
\nd

After evaluating all the integrals in eq.~\ref{eq:ap1} using the above rules for discretization the final result for $u_{i+\oh,j+\oh}^{n+1}$ is:

\bg
u_{i+\oh,j+\oh}^{n+1}=\frac{1}{4}\biglb( u^n_{i,j}+u^n_{i+1,j}+u^n_{i,j+1}+u^n_{i+1,j+1}\bigrb)+\nonumber\\
+\frac{\Delta x}{16}\partial_x\biglb( u^n_{i,j}-u^n_{i+1,j}+u^n_{i,j+1}-u^n_{i+1,j+1} \bigrb)+\frac{\Delta y}{16}\partial_y\biglb(u^n_{i,j}+u^n_{i+1,j}-u^n_{i,j+1}-u^n_{i+1,j+1} \bigrb)+\nonumber\\
+\frac{\Delta t}{2\Delta x}\Biglb[\mu\Biglb(f_{i,j}^n+f_{i,j+1}^n-f_{i+1,j}^n-f_{i+1,j+1}^n \Bigrb)+\nu\Biglb( f_{i,j}^{n+\alpha}+f_{i,j+1}^{n+\alpha}-f_{i+1,j}^{n+\alpha}-f_{i+1,j+1}^{n+\alpha} \Bigrb)\Bigrb]+\nonumber\\
+\frac{\Delta t}{2\Delta y}\Biglb[\mu\Biglb(h_{i,j}^n+h_{i+1,j}^n-h_{i,j+1}^n-h_{i+1,j+1}^n \Bigrb)+\nu\Biglb( h_{i,j}^{n+\alpha}+h_{i+1,j}^{n+\alpha}-h_{i,j+1}^{n+\alpha}-h_{i+1,j+1}^{n+\alpha} \Bigrb)\Bigrb]+\nonumber\\
+\Delta t\biglb[\frac{\xi}{4}\biglb(g_{i,j}^{n+\alpha}+g_{i+1,j}^{n+\alpha}+g_{i,j+1}^{n+\alpha}+g_{i+1,j+1}^{n+\alpha}\bigrb)+\eta g_{i+\oh,j+\oh}^{n+1}\bigrb]\nonumber\\
\label{eq:fin}
\nd

The terms at time $t^{n+\alpha}$ are taken from the solution of the predictor step:

\bg
u_{i,j}^{n+\alpha}=u_{i,j}^n-\Delta t\alpha\biglb( \partial_xf_{i,j}^n+\partial_yh_{i,j}^n-g_{i,j}^{n+\alpha}\bigrb)
\nd

As shown in \cite{Pareschi}, second order accuracy conditions give the weights used in the discretization as a function of $\alpha$.  We choose $\alpha=1/3$ with weights given by $\mu=-1/2, \nu=3/2, \xi=3/4,$ and $\eta=1/4$.  We can therefore rewrite the solution \ref{eq:fin} in operator splitting form as:

\bg
u_{i,j}^{(1)}=u_{i,j}^n-\Delta t\alpha\left( \partial_xf_{i,j}^n+\partial_yh_{i,j}^n \right) \nonumber\\
u_{i,j}^{n+\alpha}=u_{i,j}^{(1)}+\Delta t\alpha g_{i,j}^{n+\alpha} \nonumber\\
\nonumber\\
u_{i+\oh,j+\oh}^{(2)}=\frac{1}{4}\biglb( u^n_{i,j}+u^n_{i+1,j}+u^n_{i,j+1}+u^n_{i+1,j+1}\bigrb)+\nonumber\\
+\frac{\Delta x}{16}\partial_x\biglb( u^n_{i,j}-u^n_{i+1,j}+u^n_{i,j+1}-u^n_{i+1,j+1} \bigrb)+\frac{\Delta y}{16}\partial_y\biglb(u^n_{i,j}+u^n_{i+1,j}-u^n_{i,j+1}-u^n_{i+1,j+1} \bigrb)+\nonumber\\
+\frac{\Delta t}{2\Delta x}\Biglb[\mu\Biglb(f_{i,j}^n+f_{i,j+1}^n-f_{i+1,j}^n-f_{i+1,j+1}^n \Bigrb)+\nu\Biglb( f_{i,j}^{n+\alpha}+f_{i,j+1}^{n+\alpha}-f_{i+1,j}^{n+\alpha}-f_{i+1,j+1}^{n+\alpha} \Bigrb)\Bigrb]+\nonumber\\
+\frac{\Delta t}{2\Delta y}\Biglb[\mu\Biglb(h_{i,j}^n+h_{i+1,j}^n-h_{i,j+1}^n-h_{i+1,j+1}^n \Bigrb)+\nu\Biglb( h_{i,j}^{n+\alpha}+h_{i+1,j}^{n+\alpha}-h_{i,j+1}^{n+\alpha}-h_{i+1,j+1}^{n+\alpha} \Bigrb)\Bigrb]+\nonumber\\
\nonumber\\
u_{i+\oh,j+\oh}^{n+1}=u_{i+\oh,j+\oh}^{(2)}+\Delta t\biglb[\frac{\xi}{4}\biglb(g_{i,j}^{n+\alpha}+g_{i+1,j}^{n+\alpha}+g_{i,j+1}^{n+\alpha}+g_{i+1,j+1}^{n+\alpha}\bigrb)+\eta g_{i+\oh,j+\oh}^{n+1}\bigrb]\nonumber\\
\label{eq:finos}
\nd

The solution of (\ref{eq:2dstress}-\ref{eq:2dcs}) are a coupled set of seven equations of the form \ref{eq:Apde}.  At each timestep the following steps are performed.  Update the charges according to the first line of \ref{eq:finos} for $u^{(1)}$ at each point on the grid.  Then solve implicitly for $u^{n+\alpha}$ where the source terms are possibly functions of the additional six field equations.  Next update the charges according to $u^{(2)}$ at each grid point.  Do a final implicit solve for $u^{n+1}_{i+\oh,j+\oh}$ according to the final equation in \ref{eq:finos}.

\subsection{1D versus 2D}

\begin{figure}[hbtp]
  \vspace{9pt}
  \centerline{\hbox{ \hspace{0.0in}
\includegraphics[scale=0.65]{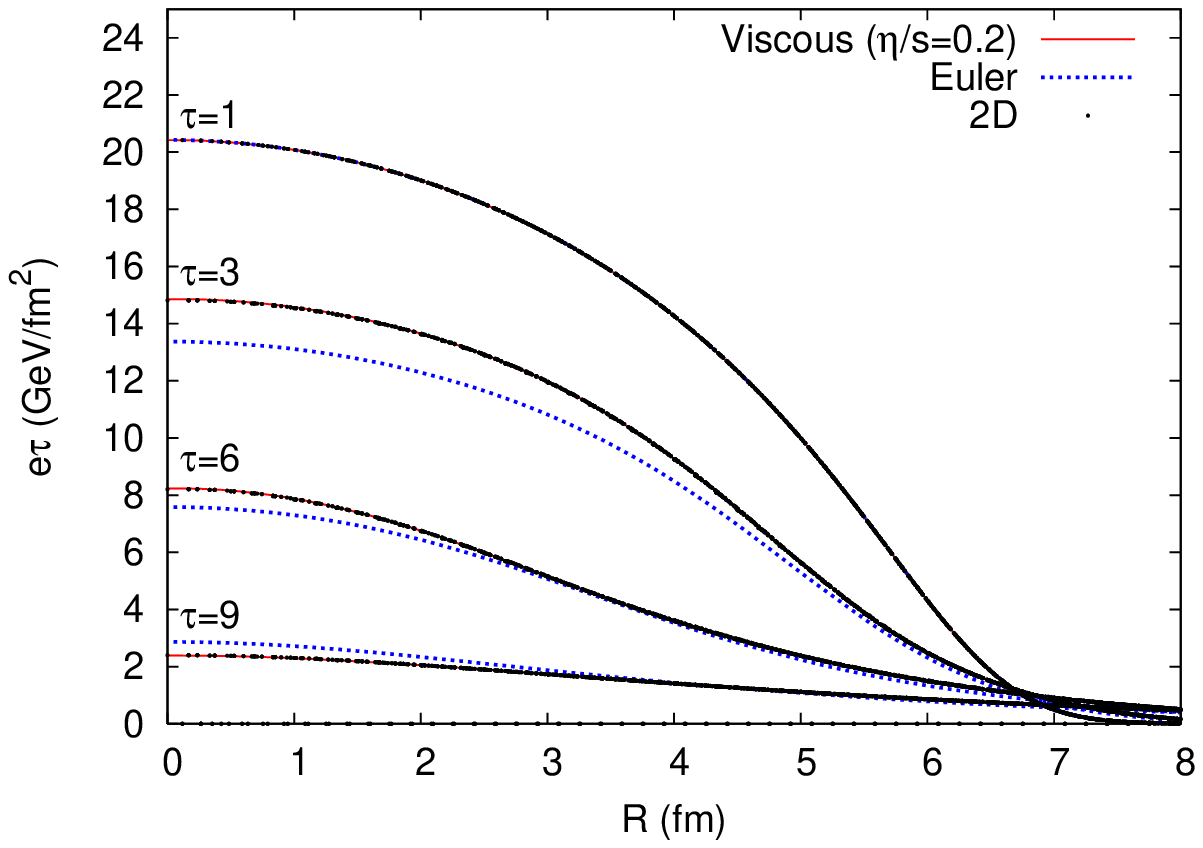}
    \hspace{0.0in}
\includegraphics[scale=0.65]{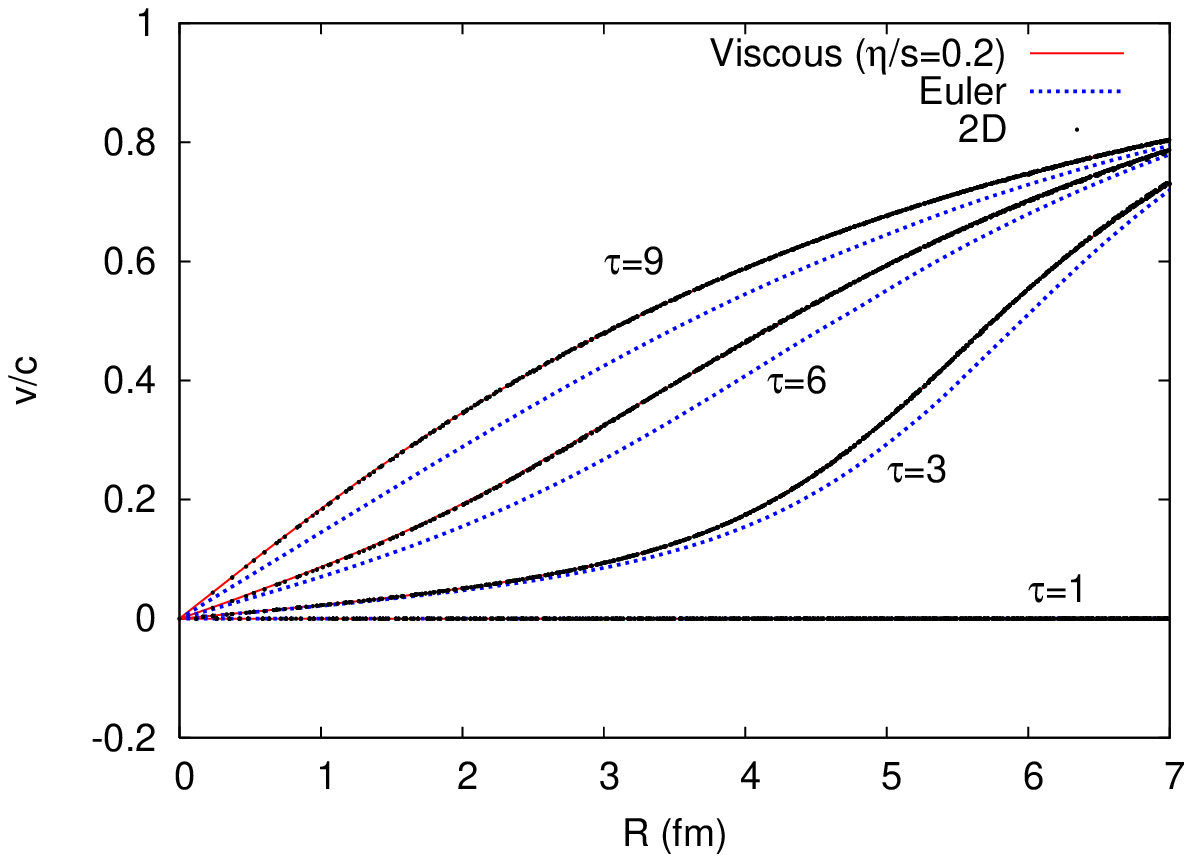}
    }
  }
\caption{Scatterplot of the energy density per unit rapidity (left) and of the transverse velocity (right).}
\label{fig:v1d2d}
\end{figure}

In order to demonstrate the robustness of the above algorithm in two dimensions we compare the results from the 2D numerical solution for central collisions with the corresponding 1D result.  In fig.~\ref{fig:v1d2d} the solid red line shows the result of the energy density per unit rapidity from the 1D case using $\eta/s=0.2$.  For reference the ideal result is shown by the dotted blue line.  The black points plotted on top of the red curve shows the corresponding result for the 2D case.  The difference between the two cases is small as expected.  The {\em scatter} in the black points gives a qualitative idea of the error due to the use of a rectangular grid.  Fig.~\ref{fig:v1d2d} shows the analogous figure for the transverse velocity.

\subsection{Gradients}
\label{sec:grad}

It was discussed in the text that Lindblom \cite{Lindblom} found an important result regarding the form of the auxillary tensor.  In a large class of causal dissipative theories the physical fluid states must relax to a state that is indistinguishable from the Navier-Stokes form.  The time scale that which this occurs in on the order of the microscopic particle interaction time.  

\begin{figure}[!ht]
  \vspace{9pt}
  \centerline{\hbox{ \hspace{0.0in} 
\includegraphics[scale=.7]{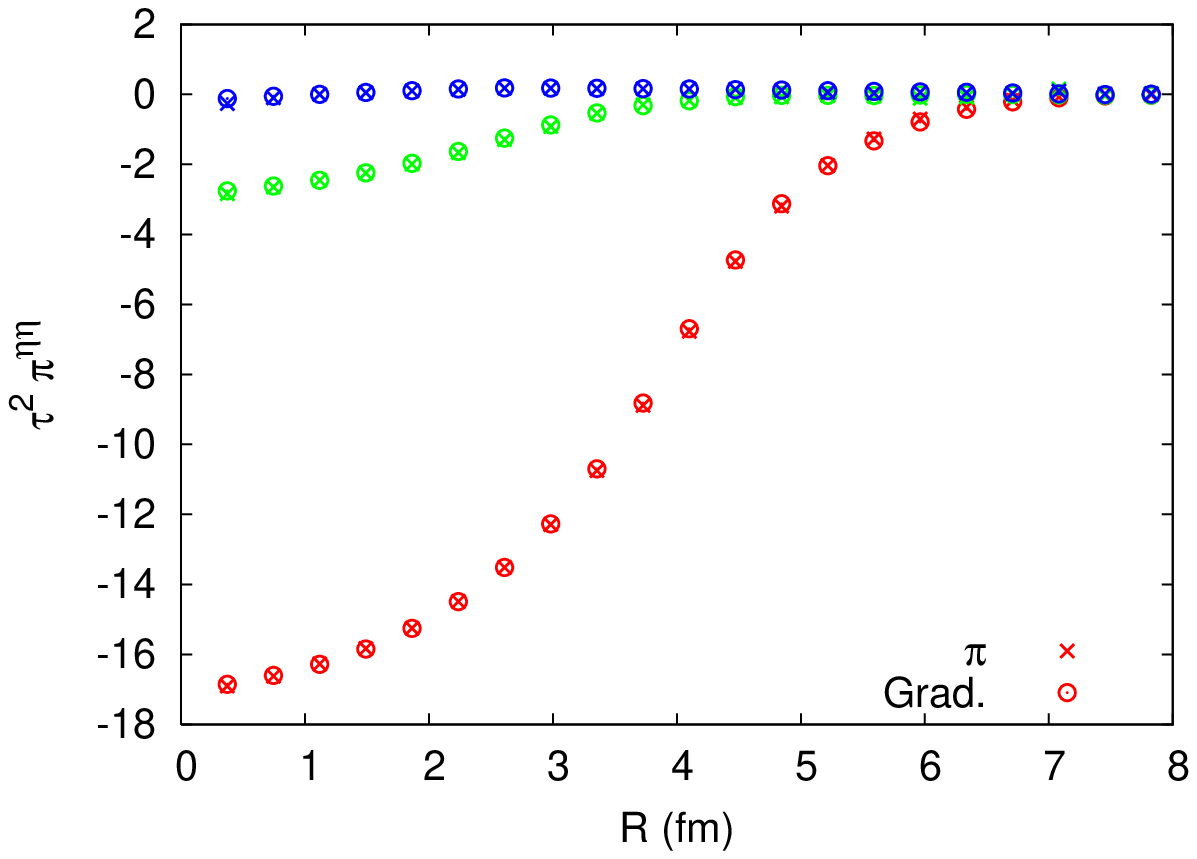}
    \hspace{0.1in}
\includegraphics[scale=.7]{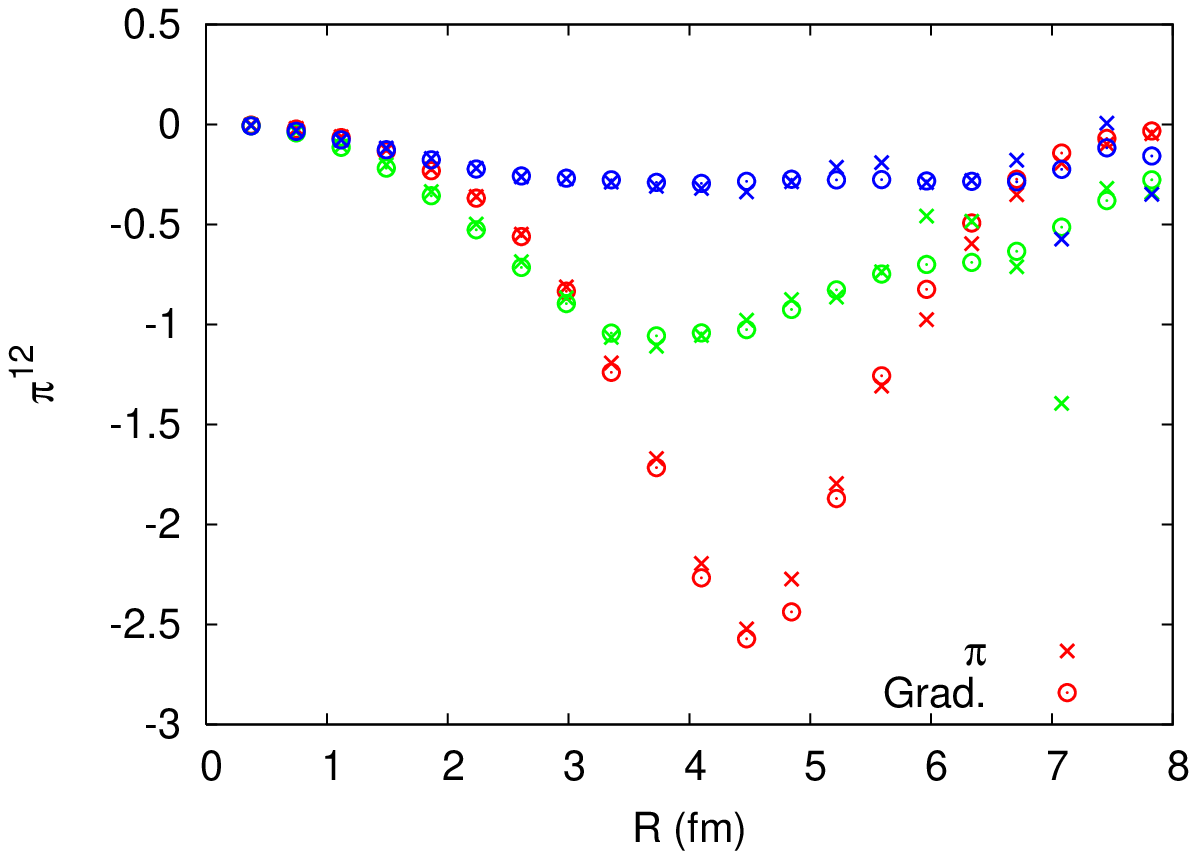}
    }
  }
  \vspace{9pt}
  \centerline{\hbox{ \hspace{0.0in}
\includegraphics[scale=.7]{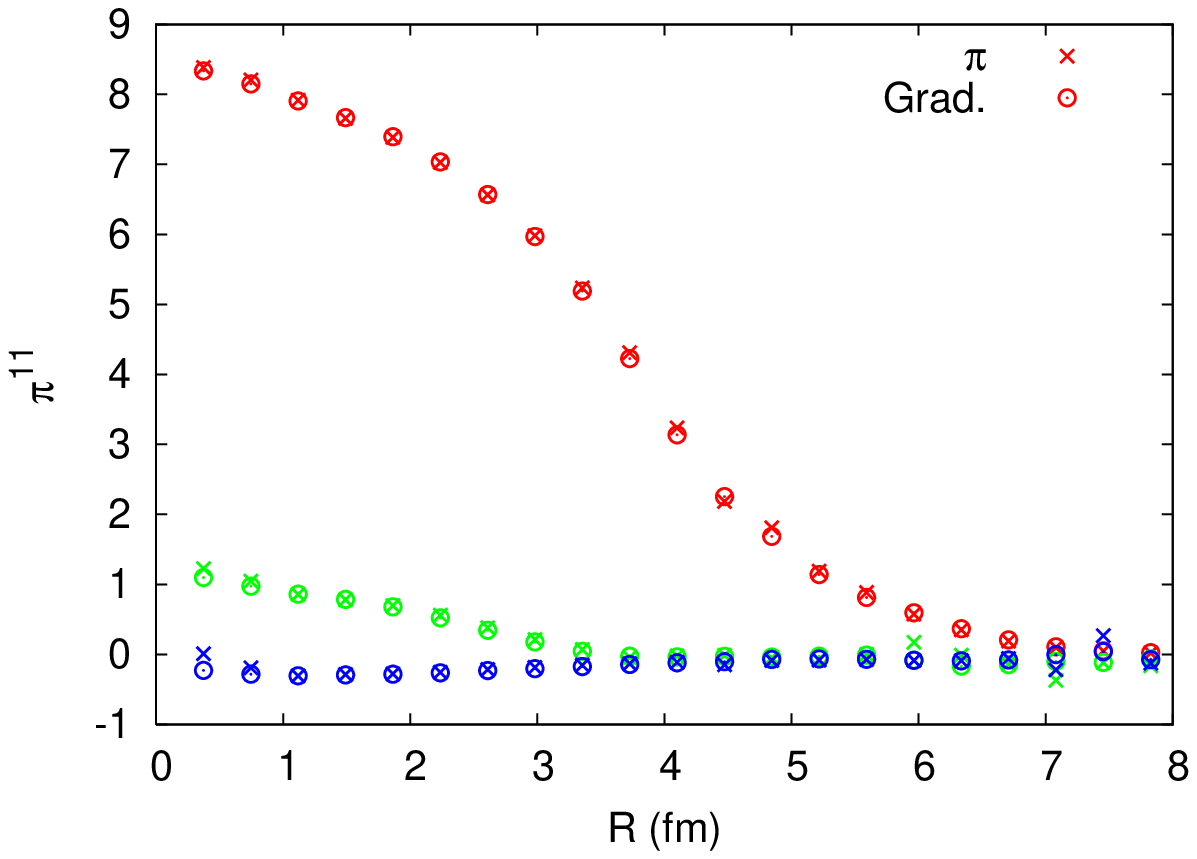}
    \hspace{0.1in}
\includegraphics[scale=.7]{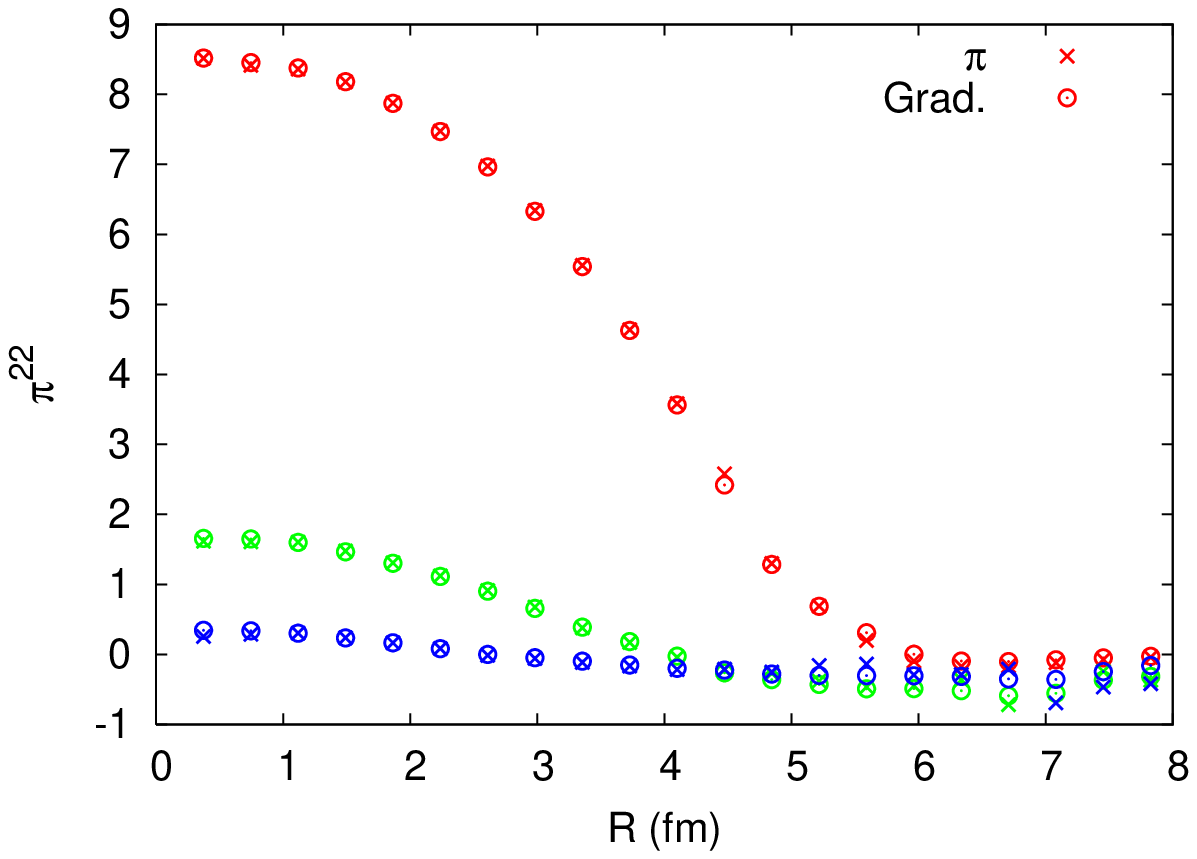}
    }
  }
  \vspace{9pt}
  \caption{(Color online) Comparison of viscous stress tensor $\pi^{\mu\nu}$ (crosses) versus its explicit calculation from the gradient of the velocity fields (open circles) for $\eta/s=10^{-6}$ for $\tau=2$ (red), $\tau=4$ (green), and $\tau=6$ (blue).  All quantities are scaled by $\eta/s$ and were generated from a slice at $30^\circ$ in the transverse plane for Au-Au collisions at b=6.5 fm.  }
  \label{grad1}
\end{figure}

\begin{figure}[!ht]
  \vspace{9pt}
  \centerline{\hbox{ \hspace{0.0in} 
\includegraphics[scale=.7]{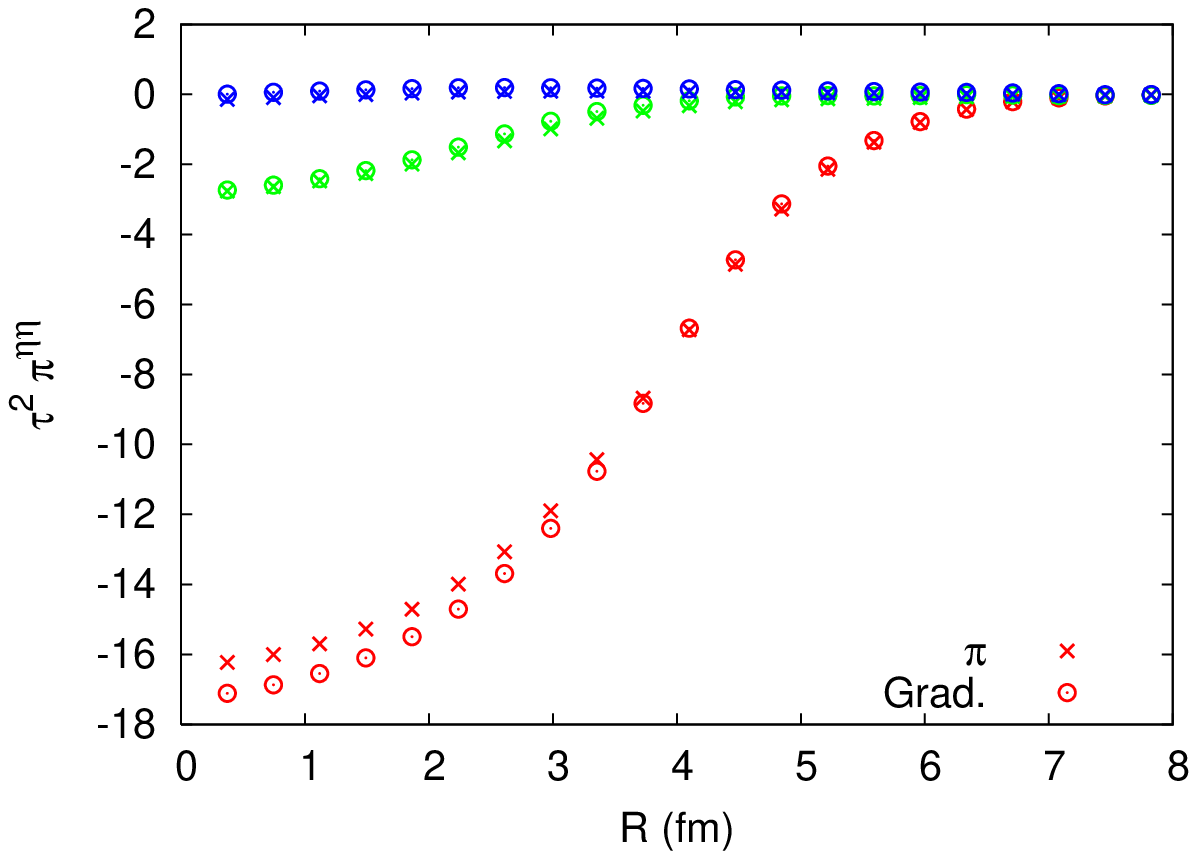}
    \hspace{0.1in}
\includegraphics[scale=.7]{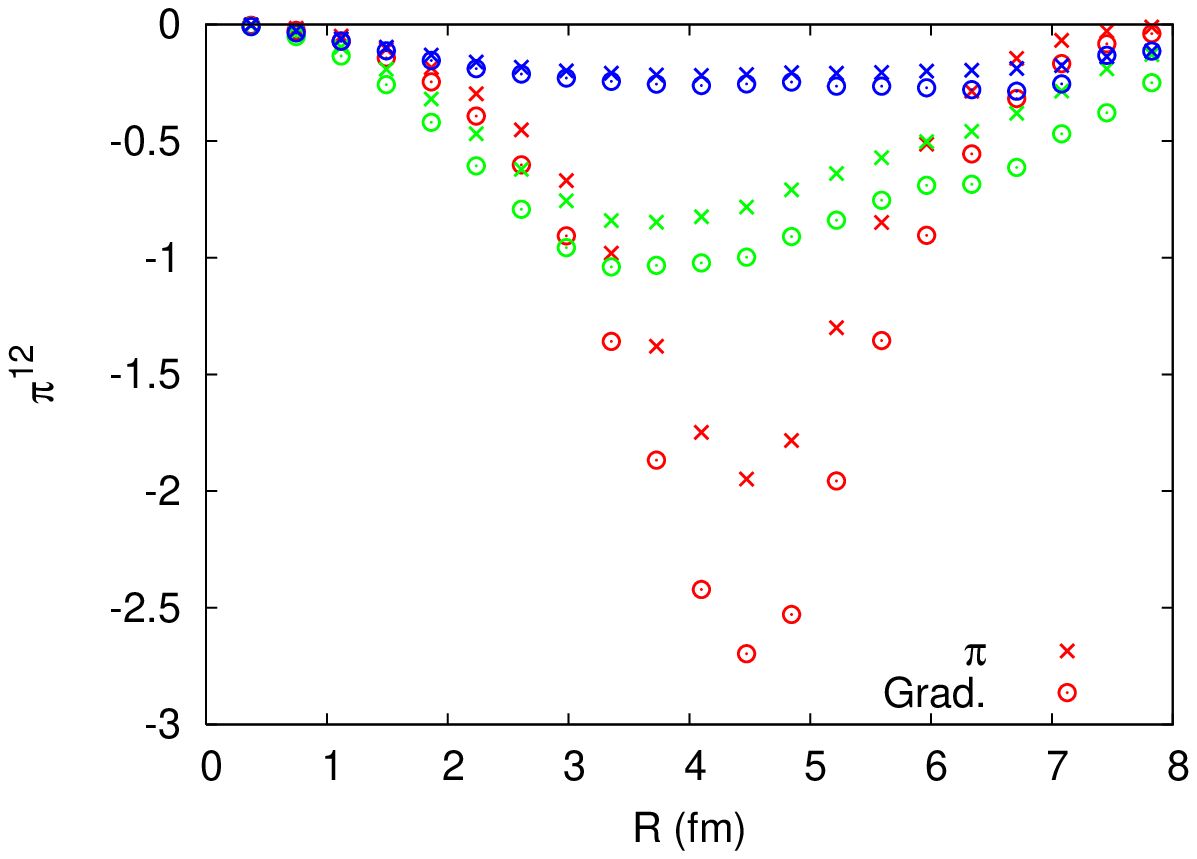}
    }
  }
  \vspace{9pt}
  \centerline{\hbox{ \hspace{0.0in}
\includegraphics[scale=.7]{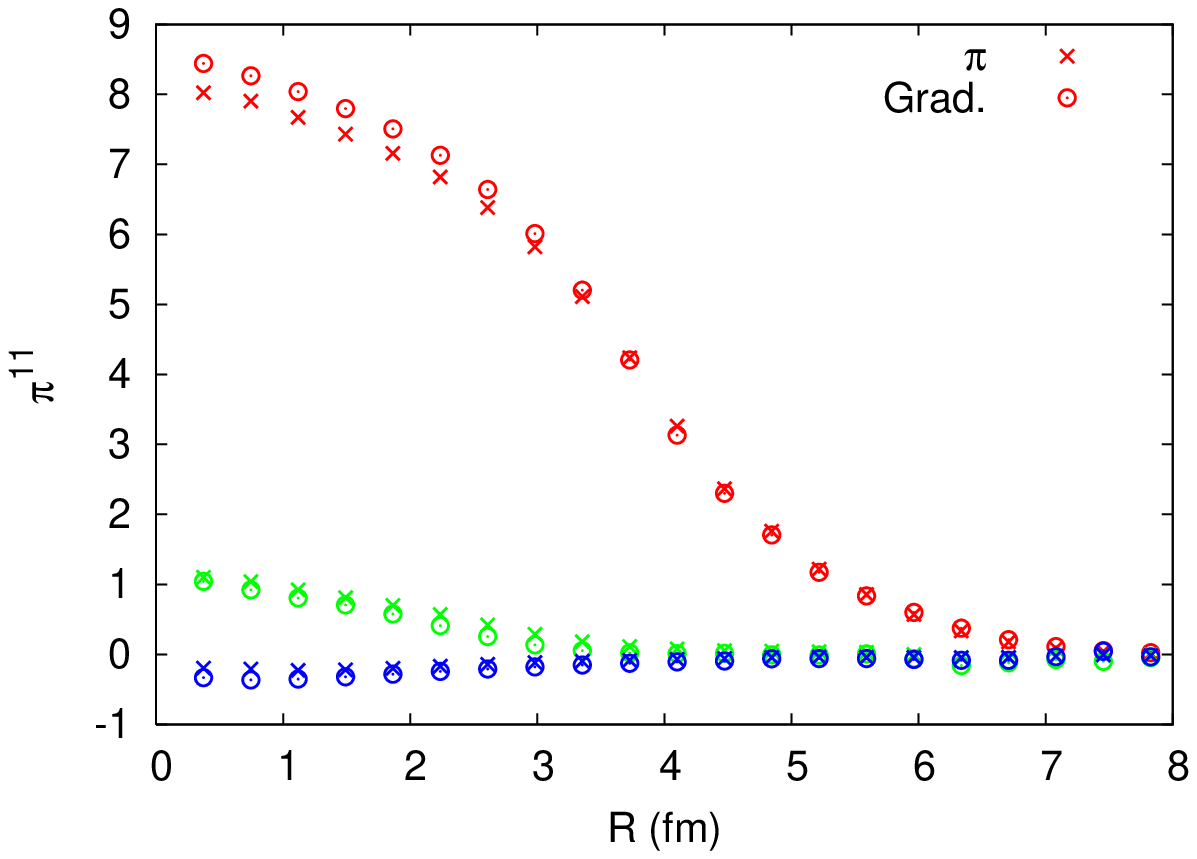}
    \hspace{0.1in}
\includegraphics[scale=.7]{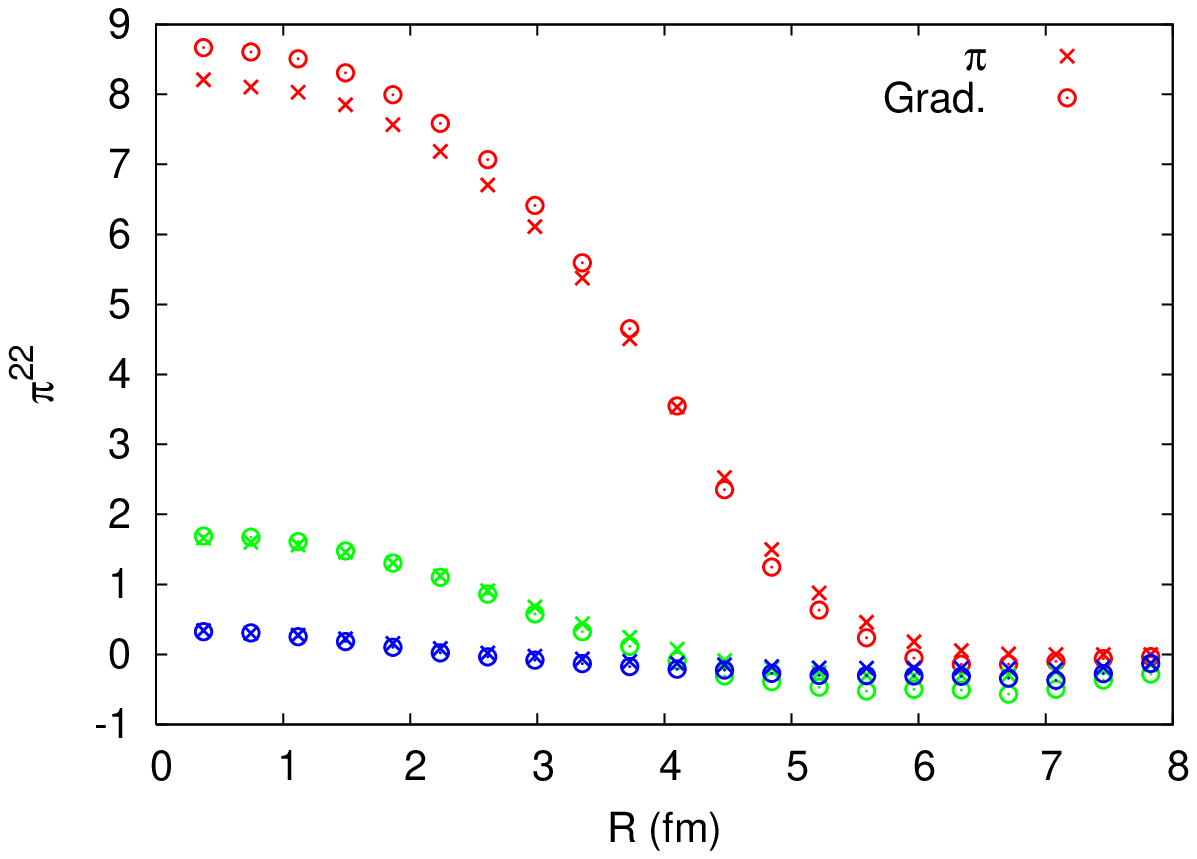}
    }
  }
  \vspace{9pt}
  \caption{(Color online) Same as fig.~\ref{grad1} except for $\eta/s=0.05$.}
  \label{grad2}
\end{figure}

\begin{figure}[!ht]
  \vspace{9pt}
  \centerline{\hbox{ \hspace{0.0in} 
\includegraphics[scale=.7]{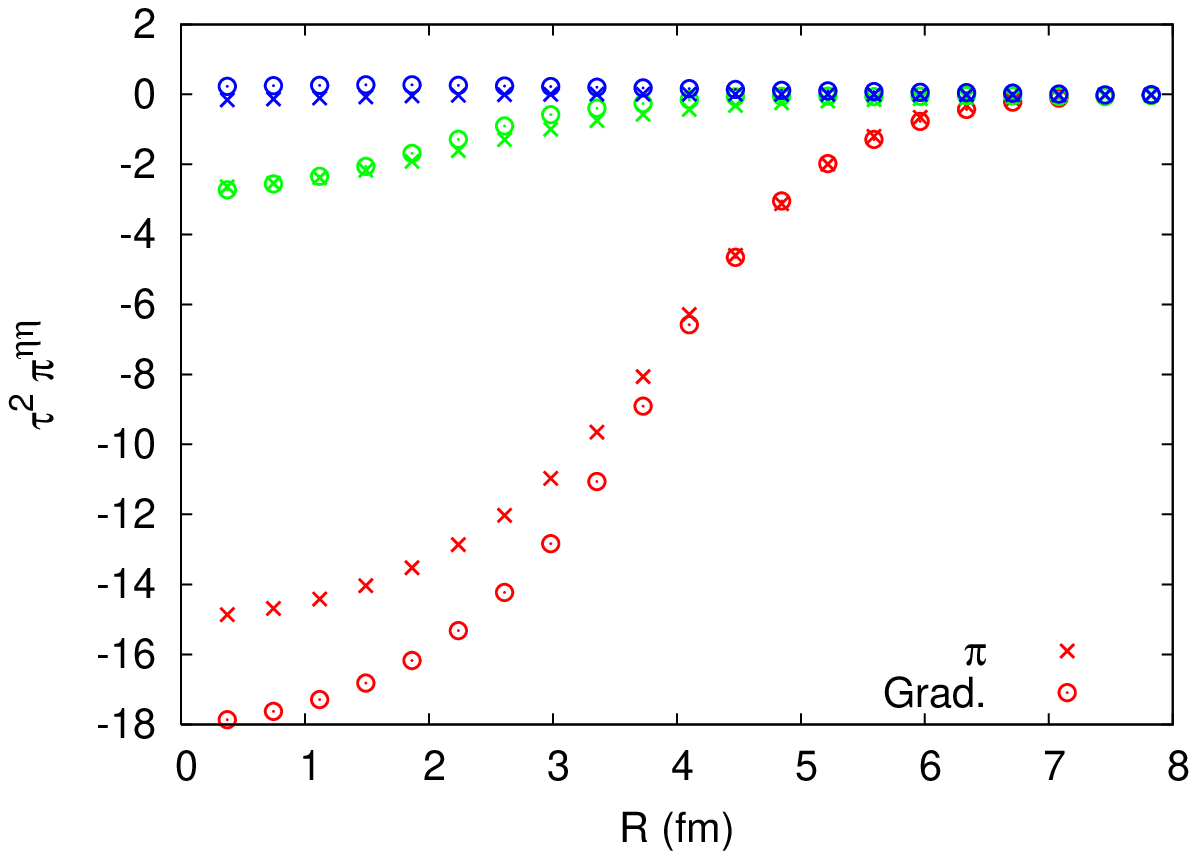}
    \hspace{0.1in}
\includegraphics[scale=.7]{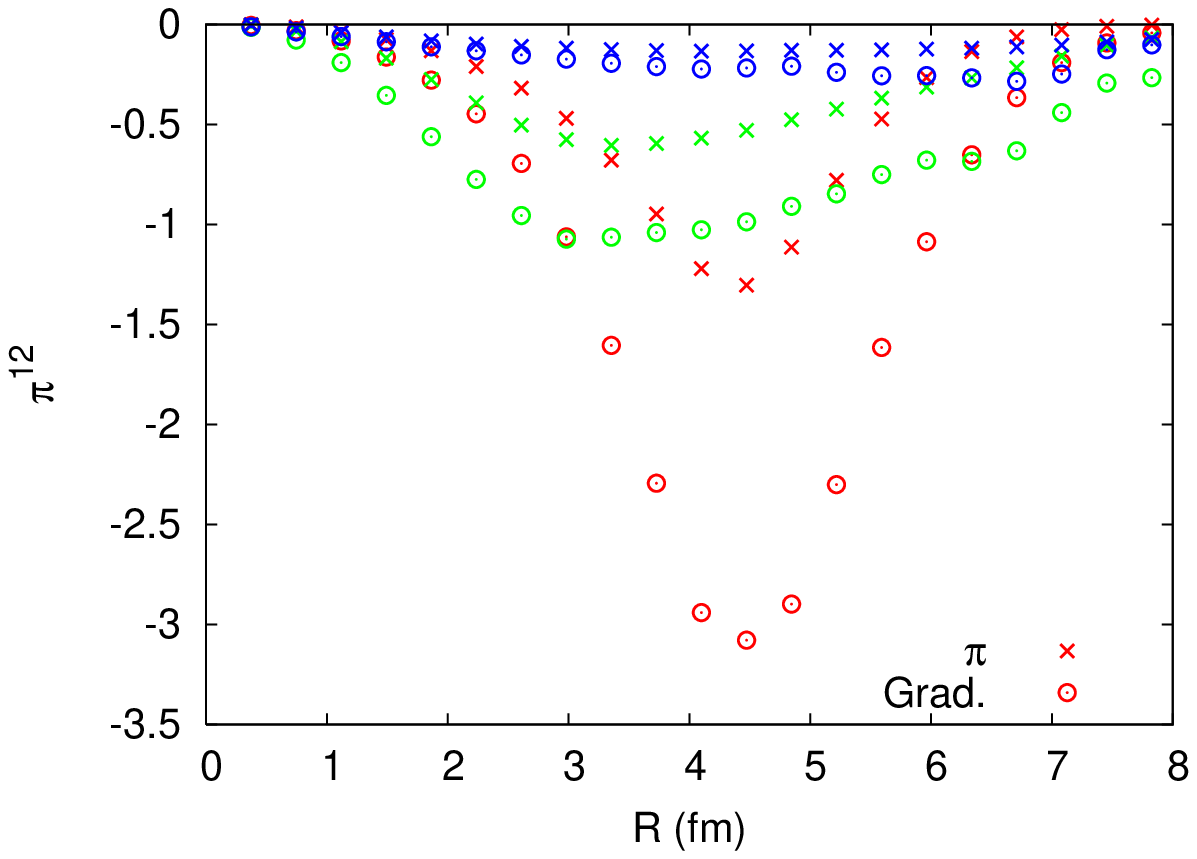}
    }
  }
  \vspace{9pt}
  \centerline{\hbox{ \hspace{0.0in}
\includegraphics[scale=.7]{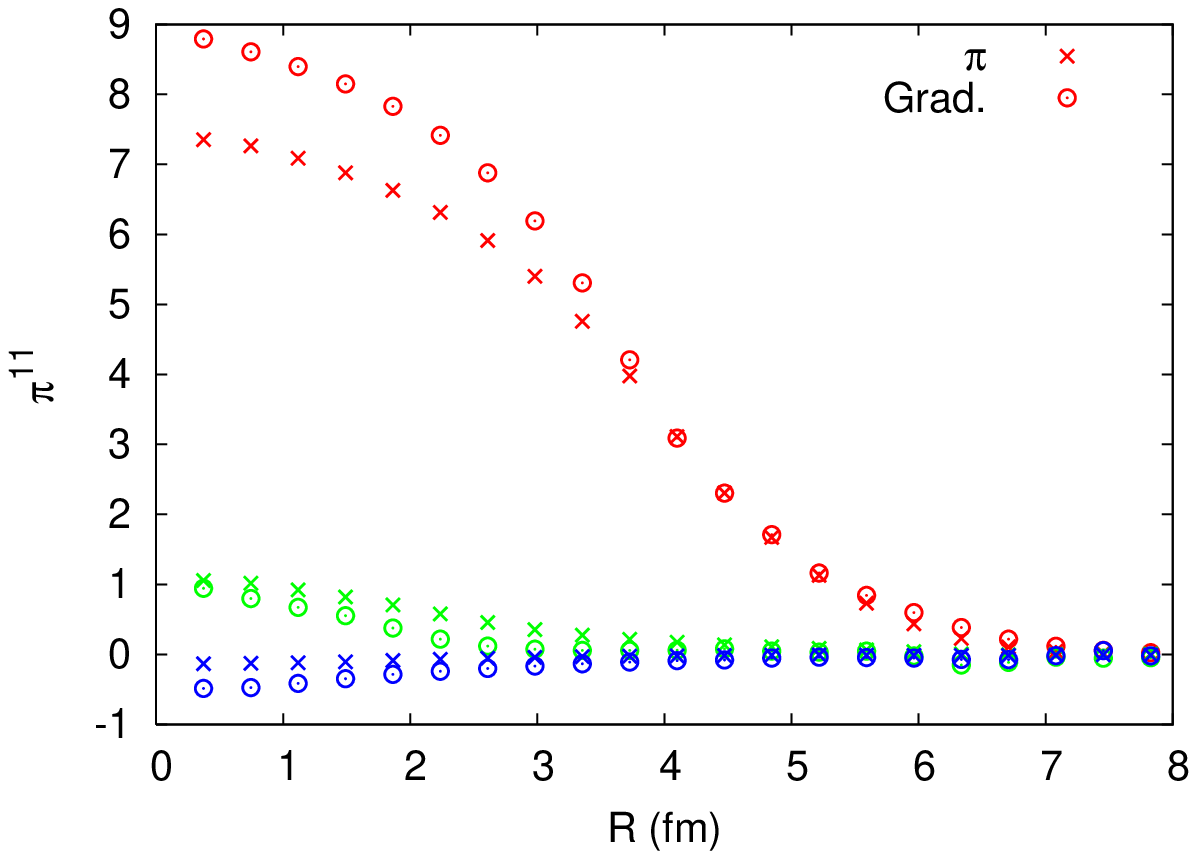}
    \hspace{0.1in}
\includegraphics[scale=.7]{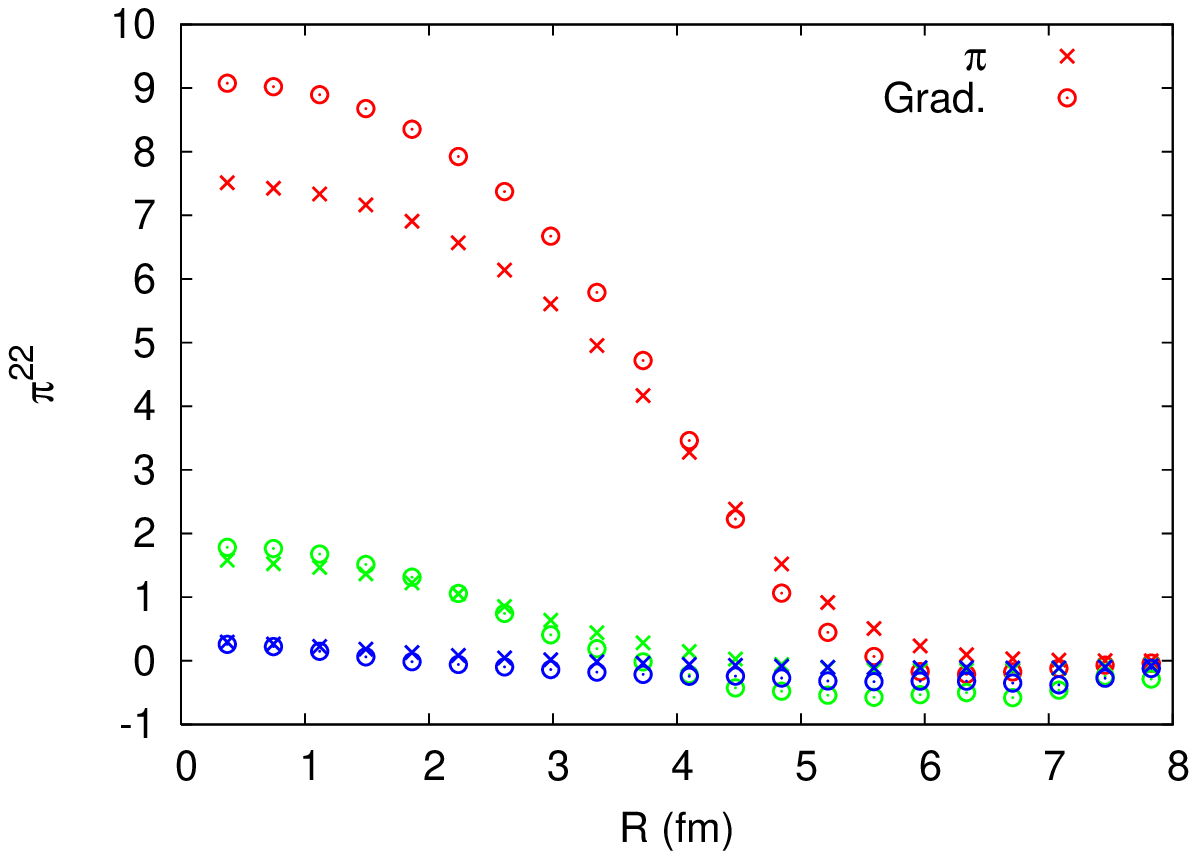}
    }
  }
  \vspace{9pt}
  \caption{(Color online) Same as fig.~\ref{grad1} except for $\eta/s=0.2$.}
  \label{grad3}
\end{figure}

We therefore should check that the viscous stress tensor $\pi^{\mu\nu}$ as computed from the auxiliary tensor $c^{\mu\nu}$ agrees with the stress tensor computed from the gradients of the velocity field.  This is shown for various components of $\pi^{\mu\nu}$ from simulations with viscosity of $\eta/s=10^{-6}, 0.05, 0.2$ in figs.~\ref{grad1}-\ref{grad3}. 

\end{document}